\documentclass[twocolumn]{aastex631}
\usepackage{amsmath,braket,bm,ulem}
\usepackage{natbib}
\usepackage{soul}
\usepackage{hhline}
\definecolor{darkgreen}{rgb}{0,0.60,0}

\definecolor{cyan}{rgb}{0,0.8,0.8}

\usepackage[nonumberlist,nosuper]{glossaries}
\setacronymstyle{long-short}
\newacronym{1d}{1D}{one-dimensional}
\newacronym{3d}{3D}{three-dimensional}
\newacronym{aa}{AA}{angle-averaged}
\newacronym{ali}{ALI}{accelerated $\Lambda$-iteration}
\newacronym{crd}{CRD}{complete frequency redistribution}
\newacronym{fs}{FS}{fine-structure}
\newacronym{hfs}{HFS}{hyperfine structure}
\newacronym{hf}{HF}{hyperfine}
\newacronym{ipb}{IPB}{incomplete Paschen-Back}
\newacronym{los}{LOS}{lines of sight}
\newacronym{lte}{LTE}{local thermodynamical equilibrium}
\newacronym{prd}{PRD}{partial frequency redistribution}
\newacronym{rt}{RT}{radiative transfer}
\newacronym{rte}{RTE}{radiation transfer equations}
\newacronym{see}{SEE}{statistical equilibrium equations}
\newacronym{wfa}{WFA}{weak field approximation}
\newacronym{uv}{UV}{ultraviolet}

\begin{document} 
\title{The polarization of the solar Ba~{\sc ii} D$_1$ line with partial frequency redistribution \\ and its magnetic sensitivity}
\author[0000-0001-9095-9685]{Ernest Alsina Ballester}
\affiliation{Instituto de Astrof\'{i}sica de Canarias, E-38205 La Laguna, Tenerife, Spain}
\affiliation{Departamento de Astrof\'{i}sica, Universidad de La Laguna, E-38206 La Laguna, Tenerife, Spain}

\author[0000-0003-1465-5692]{Tanaus\'u del Pino Alem\'an}
\affiliation{Instituto de Astrof\'{i}sica de Canarias, E-38205 La Laguna, Tenerife, Spain}
\affiliation{Departamento de Astrof\'{i}sica, Universidad de La Laguna, E-38206 La Laguna, Tenerife, Spain}

\author[0000-0001-5131-4139]{Javier Trujillo Bueno}
\affiliation{Instituto de Astrof\'{i}sica de Canarias, E-38205 La Laguna, Tenerife, Spain}
\affiliation{Departamento de Astrof\'{i}sica, Universidad de La Laguna, E-38206 La Laguna, Tenerife, Spain}
\affiliation{Consejo Superior de Investigaciones Científicas, Spain}

\begin{abstract}
We investigate the main physical mechanisms that shape the intensity and polarization of the 
Ba~{\sc{ii}} D$_1$ line {at $4934$~\AA } via radiative transfer numerical experiments. 
We focus especially on the scattering linear polarization arising from the spectral structure of 
the anisotropic radiation in the wavelength interval spanned by the line's \gls*{hfs} components in the odd isotopes of barium. 
After verifying that the presence of the low-energy metastable levels only impacts the amplitude, 
but not the shape, of the D$_1$ linear 
polarization, we relied on a two-term atomic model that neglects such metastable levels but includes \gls*{hfs}. 
The D$_1$ fractional linear polarization shows a very small variation with the choice of atmospheric model, enhancing 
its suitability for solar magnetic field diagnostics. Tangled 
magnetic fields with strengths of tens of gauss 
reduce the linear polarization and saturation is reached at {roughly $300$~G.}
Deterministic inclined magnetic fields produce a $U/I$ profile and, if they have a significant longitudinal component, a $V/I$ 
profile, whose modeling requires accounting for \gls*{hfs} and the Paschen-Back effect. 
Because of the overlap between \gls*{hfs} components, the 
magnetograph formula cannot be applied to infer the longitudinal magnetic field.  
Accurately modeling the D$_1$ intensity and polarization requires an atomic system that 
includes the metastable levels and the \gls*{hfs}, 
the detailed spectral structure of the radiation field, {the incomplete Paschen-Back regime for magnetic fields}, 
and an accurate treatment of collisions. 
\end{abstract}


\section{Introduction}
\label{sec::intro}
High-precision spectropolarimetric observations in quiet regions close to the solar limb reveal a wealth of 
linearly polarized features in spectral lines, known as the second solar spectrum 
\citep[e.g.,][]{Ivanov91,StenfloKeller97}. Such polarization signals arise from the scattering of anisotropic 
radiation (i.e., scattering polarization) within the solar atmosphere. 
Through measurements of the scattering polarization, valuable information on the properties of the solar atmosphere can 
be obtained. Indeed, the scattering polarization in spectral lines is generally sensitive to the magnetic field via the
Hanle effect \citep[e.g.,][]{BStenflo94,BLandiLandolfi04}, which enables practical diagnostics of solar magnetic fields, 
especially in the upper chromosphere and transition region 
as well as {in} solar prominences or spicules \citep[e.g., the review by][]{TrujilloBuenodelPinoAleman22}, or 
at sub-resolution scales \citep[e.g.,][]{TrujilloBueno+04} which cannot easily be accessed with more widespread 
techniques {such as those based on the Zeeman effect}. 

The D lines of Ba~{\sc{ii}} encode valuable information on the atmospheric properties of the lower 
solar chromosphere (see Appendix~\ref{sec-app::Formation} for the formation height of D$_1$). 
Over the last two decades, a considerable volume of spectropolarimetric observations of the 
Ba~{\sc{ii}} D$_2$ lines has been acquired 
\citep[e.g.,][]{Faurobert+09,LopezAriste+09,Ramelli+09} and several theoretical investigations on its large 
scattering polarization {signal} 
and {its} sensitivity to the Hanle effect have been carried out \citep[e.g.,][]{Belluzzi+07,Faurobert+09,Smitha+13}.  
The first observations of the linear polarization of the solar Ba~{\sc{ii}} D$_1$ line revealed two positive peaks, 
with the blue (red) one above (below) the continuum level (see \citealt{StenfloKeller97}; {also \citealt{Stenflo+2000}}). 
The fact that, in quiet regions, this line and the D$_1$ line of Na~{\sc{i}} did not simply present a depolarized 
feature was regarded as surprising, because the upper and lower \gls*{fs} levels of these lines have angular 
momentum $J = 1/2$ (i.e., they are intrinsically unpolarizable). 
A compelling explanation for these features was eventually put forward by \cite{BelluzziTrujilloBueno13}, 
which relied on the \gls*{hfs} present in all sodium isotopes and in the odd barium isotopes 
($^{135}$Ba and $^{137}$Ba, which represent $18\%$ of the total). 
Their modeling took into account the frequency correlations between the incident and scattered radiation, that is, 
including \gls*{prd} effects. Thus, they could account for the spectral structure of the anisotropic radiation field 
over the wavelength intervals spanned by these lines' \gls*{hfs} components and showed that this gives rise 
to linear polarization signals comparable to the observed ones. 
Subsequently, \cite{AlsinaBallester+21} modeled the Na~{\sc{i}} D lines accounting for this spectral structure and additionally 
considering the frequency redistribution effects of elastic collisions and magnetic fields. 
Their calculations 
showed that scattering polarization signals of substantial amplitude can be produced in the intrinsically unpolarizable D$_1$ lines, 
even in the presence of gauss-strength magnetic fields typical of the quiet Sun. 
Moreover, the D$_1$ line was also shown to be sensitive to such magnetic fields, adding to its diagnostic interest.  

In this work, we carry out an analogous investigation for the Ba~{\sc{ii}} D$_1$ line in which, for the first time, we jointly 
account for scattering polarization with \gls*{prd} effects, the \gls*{hfs} {of the atom}, quantum interference between atomic 
states belonging to the same term, and magnetic fields of arbitrary strength. 
The D lines of both Ba~{\sc{ii}} and Na~{\sc{i}} originate from resonance transitions between 
a lower $s$ term and an upper $p$ term. The upper \gls*{fs} levels of the D$_1$ and D$_2$ lines have 
$J = 1/2$ and $3/2$, respectively, and the ground term has a single \gls*{fs} level with $J = 1/2$. Nevertheless, 
there are important differences between the atomic structure of the two atomic species. 
The separation between the upper levels of the Ba~{\sc{ii}} D$_1$ line at $4934$~\AA\ and of the D$_2$ line at $4554$~\AA\ 
is much larger than the corresponding separation for the case of the Na~{\sc{i}} atom. 
Moreover, for the isotopes with nuclear spin, the \gls*{hfs} splitting of the upper and lower levels of Ba~{\sc{ii}} is 
roughly one order of magnitude larger than that of the corresponding levels of Na~{\sc{i}}. 
It is also noteworthy that the Ba~{\sc{ii}} {system} has a metastable term $5d \, ^{2}\mathrm{D}$, whose two 
\gls*{fs} levels have significantly lower energies than the upper term of the D lines. 
To our knowledge, the impact of the metastable levels on the D$_1$ intensity and polarization has not been 
studied to date. 

This paper is organized as follows. In Section~\ref{sec::problem}, we introduce the basic assumptions and the
most general atomic model considered in this work. In Section~\ref{sec-sub::metastable}, we theoretically study the impact on the 
D$_1$ line of the metastable levels. Building on these results, Section~\ref{sec::numerical} is focused on a series of numerical 
experiments on the Ba~{\sc{ii}} D$_1$ line, in which the metastable levels are neglected. We study {how the intensity and polarization 
patterns of the line are impacted by the} \gls*{hfs} splitting and the quantum interference between \gls*{hfs} and \gls*{fs} levels, 
as well as the sensitivity of the line to different atmospheric models and to magnetic fields, both {isotropically distributed} and deterministic. 
Conclusions are outlined in Section~\ref{sec::conclusions}. 
Information about the atomic quantities used in this work and additional figures, including those pertaining to the Ba~{\sc{ii}} 
D$_2$ line, can be found in the appendices. 

\section{Formulation of the problem}
\label{sec::problem}
Our theoretical investigation {aims at highlighting}  the impact of various physical mechanisms on the intensity and polarization 
of the Ba~{\sc{ii}} D$_1$ line. {Such investigations are based on} a series of spectral syntheses, obtained through the numerical solution 
of the \gls*{rt} problem out of \gls*{lte} conditions. 
{We account for \gls*{prd} effects, in order to suitably account for} 
{the spectral structure of the incident radiation field, which can  
introduce a linear polarization signal in the D$_1$ line as explained in} \cite{BelluzziTrujilloBueno13}.  
For the sake of reducing computational cost, we decouple the angular and 
frequency dependence introduced by the Doppler effect in scattering processes by making the 
\gls*{aa} approximation \citep{ReesSaliba82}, except where otherwise noted. 
We {considered} \gls*{1d} semiempirical atmospheric models, namely those introduced in \cite{Fontenla+93} --- hereafter FAL models --- 
and the M$_{\mbox{\scriptsize{CO}}}$ model of \cite{Avrett95} --- hereafter FAL-X. In particular, we considered the FAL-C model except where otherwise noted. 
The \gls*{los} for which we show the synthetic profiles are given by $\mu = \cos\theta$, where $\theta$ is the heliocentric angle. 
In order to consider scattering polarization profiles of substantial amplitude, we take $\mu = 0.1$ except where otherwise noted. 
In all cases, we take the reference direction for positive $Q$ to be parallel to the nearest limb. 

\begin{figure}[!h]
 \centering
 \includegraphics[width = 0.485\textwidth]{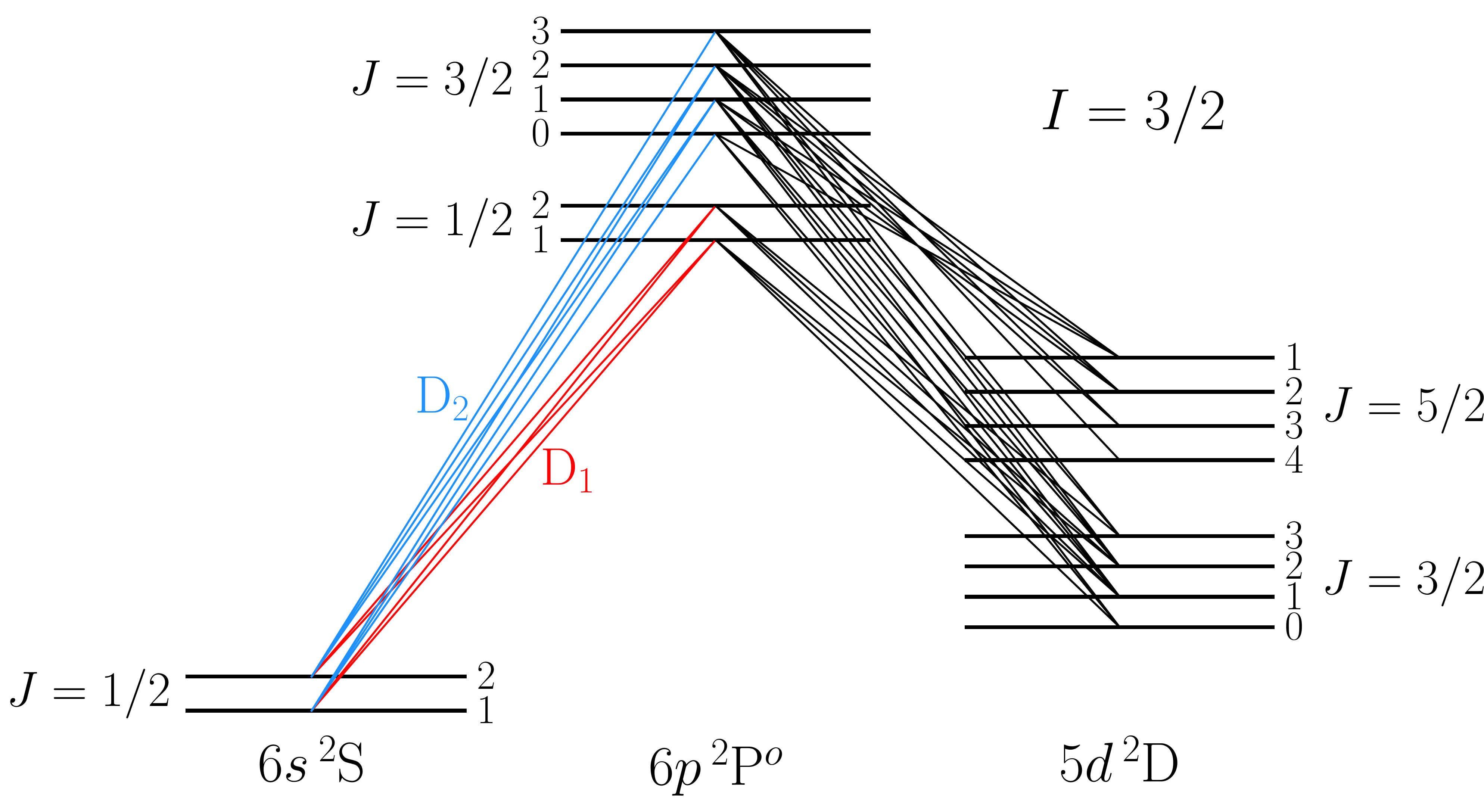}
 \caption{Grotrian diagram for the most general Ba~{\sc{ii}} model considered in this work, which includes three terms:  
 $6s \,^{2}\mathrm{S}$ (ground term), $5d \,^{2}\mathrm{D}$ (metastable term), 
 and $6p \,^{2}\mathrm{P}^{\mathrm{o}}$ (upper term for the D lines). The latter two consist of two \gls*{fs} levels each. 
 The figure also displays the \gls*{hfs} for {the case of} 
the odd isotopes (energies not to scale). The solid lines show the {permitted} radiative transitions 
that couple the atomic states of the system. 
The red and blue lines indicate the transitions pertaining to the D$_1$ and D$_2$ lines, respectively. }
  \label{fig-app::FApp2-Grotrian}
\end{figure} 
Except where otherwise noted, we considered the seven stable isotopes of barium throughout this work. Of these, only the two odd 
isotopes ($^{135}$Ba and $^{137}$Ba) have a nonzero nuclear spin with $I = 3/2$ and thus \gls*{hfs}. Relevant atomic quantities, 
including the isotopic abundances and shifts and the \gls*{hfs} coefficients can be found in Appendix~\ref{sec-app::AtomicQuantTables}. 
Outside sunspots, Ba~{\sc{i}} is a minority species and we thus consider all barium atoms to be in the Ba~{\sc{ii}} and Ba~{\sc{iii}} stages. 
In the most general case, we consider five \gls*{fs} levels of Ba~{\sc{ii}}, namely $6s\,^2\mathrm{S}_{1/2}$ (the ground level), 
$5d\,^2\mathrm{D}_{3/2}$ and $5d\,^2\mathrm{D}_{5/2}$ (the metastable levels), and $6p\,^{2}\mathrm{P}^{\mathrm{o}}_{1/2}$ and 
$6p\,^{2}\mathrm{P}^{\mathrm{o}}_{3/2}$ (the upper levels of D$_1$ and D$_2$, respectively). 
The other levels in this ionization stage, which have at least twice the energy of the $6p$ term, are not considered in 
this work. The Grotrian diagram for this atomic model, including {the} \gls*{hfs}, can be found in Figure~\ref{fig-app::FApp2-Grotrian}. 
Each \gls*{hfs} level is indicated by {its} corresponding quantum number $F$. We observe that the energies of the \gls*{hfs} levels of 
the $5d\,^2\mathrm{D}_{5/2}$ metastable level decrease with $F$, as a consequence of its negative magnetic dipole \gls*{hfs} coefficient 
$\mathcal{A}$ (see Appendix~\ref{sec-app::AtomicQuantTables}).  
The solid lines connecting the various $F$ levels indicate the permitted radiative transitions between them, with red and blue lines 
pertaining to the D$_1$ and D$_2$ lines, respectively. We only account for the ground level of Ba~{\sc{iii}}, which we consider suitable 
for determining the ionization balance. 

The results of the \gls*{rt} calculations and the corresponding analysis are presented in the following two sections. 
In Section~\ref{sec-sub::metastable}, we study the impact of the metastable levels on the D$_1$ linear polarization in the absence of magnetic fields, 
using the HanleRT \citep{delPinoAleman+16,delPinoAleman+20} synthesis code. 
After establishing that the metastable levels modify the amplitude but not the shape of the D$_1$ scattering polarization, in Section~\ref{sec::numerical} 
we study the impact of the \gls*{hfs}, the atmospheric model, and the magnetic fields on the polarization patterns of the D$_1$ line, 
neglecting the metastable levels. Such numerical investigations were carried out using the \gls*{rt} code for a two-term model introduced in \cite{AlsinaBallester+22}. 

\section{The impact of the {5$d$} metastable levels in the zero-field case} 
\label{sec-sub::metastable}
At present, no \gls*{rt} code exists that can simultaneously account for \gls*{prd} effects, the five above-mentioned atomic levels 
of Ba~{\sc{ii}}, {the} quantum interference between levels belonging to the same term, the \gls*{hfs}, and magnetic fields in the 
\gls*{ipb} effect regime. 
However, we can still gain valuable insights into the physics that shape the intensity and polarization patterns of the 
Ba~{\sc{ii}} D lines by employing different numerical approaches that can each account for most of the aforementioned phenomena. 

In the present subsection, we made use of the HanleRT numerical code for the synthesis of the intensity and polarization of the D lines. 
{The HanleRT code accounts for scattering processes with \gls*{prd} effects} following the formalism introduced by 
\citet[][]{Casini+14,Casini+17a,Casini+17b} and, in its present version, can consider multi-term
\footnote{In contrast to a multi-level atomic model, a multi-term model accounts for the quantum interference between different \gls*{fs} levels of the 
same term {\cite[see, e.g., Section 7.5 of][]{BLandiLandolfi04}}.} atomic systems without \gls*{hfs}. 
A multi-level modeling that includes the \gls*{hfs} and the quantum interference between the $F$ levels pertaining to a given $J$ 
level can be achieved with HanleRT by making the formal substitutions $S\rightarrow I$, $J\rightarrow F$, and $L\rightarrow J$, considering the 
\gls*{hfs} splitting introduced in Appendix~\ref{sec-app::AtomicQuantTables}. This treatment neglects quantum interference between \gls*{fs} 
levels which, as confirmed in \ref{sec-sub::num-hfs}, is a reasonable assumption. This approach is otherwise correct in the absence of magnetic 
fields. However, in the presence of magnetic fields, the same substitution 
leads to an incorrect expression of the magnetic Hamiltonian \citep[{e.g.,}][]{Janett+23}. Thus, the calculations with HanleRT presented
in this work are restricted to the nonmagnetic case.  

For the calculations carried out with HanleRT, we considered only the five most abundant isotopes; we 
neglected the contribution from the two least abundant stable isotopes because data on the isotopic shifts 
of their corresponding metastable levels is, as far as we are aware, not presently available. 
The abundance of the remaining five isotopes was adjusted accordingly. We expect the error incurred to be negligible, because of the low 
abundance of the omitted isotopes which, moreover, have no nuclear spin. 

The population of the ground term, $\mathcal{N}_\ell$, {was} kept fixed during the iterative solution
of the non-\gls*{lte} \gls*{rt} problem, while letting the overall population of the Ba~{\sc{ii}} 
levels evolve freely. 
The reason for this is two-fold. First, the population and ionization balance is calculated considering only the $^{138}$Ba isotope. 
This population is then distributed among the isotopes according to their abundances and, for those with \gls*{hfs}, the populations 
in a given \gls*{fs} level are distributed among the $F$ levels according to their statistical weight. Secondly, the metastable levels are 
critical for the population balance of the atom. If we were to completely fix the populations, we would be prescribing the populations 
in the $F$ levels whereas, if we were to leave them completely free (thus ensuring mass conservation), we would not be able to 
analyze the actual impact on the linear polarization, because the population balance will significantly change the intensity profile.  
{W}e consider this a reasonable {approximation} because the population of the lower level is much larger than that of the other levels 
of the system. 

\begin{figure}[!h]
\centering
\includegraphics[width = 0.485\textwidth]{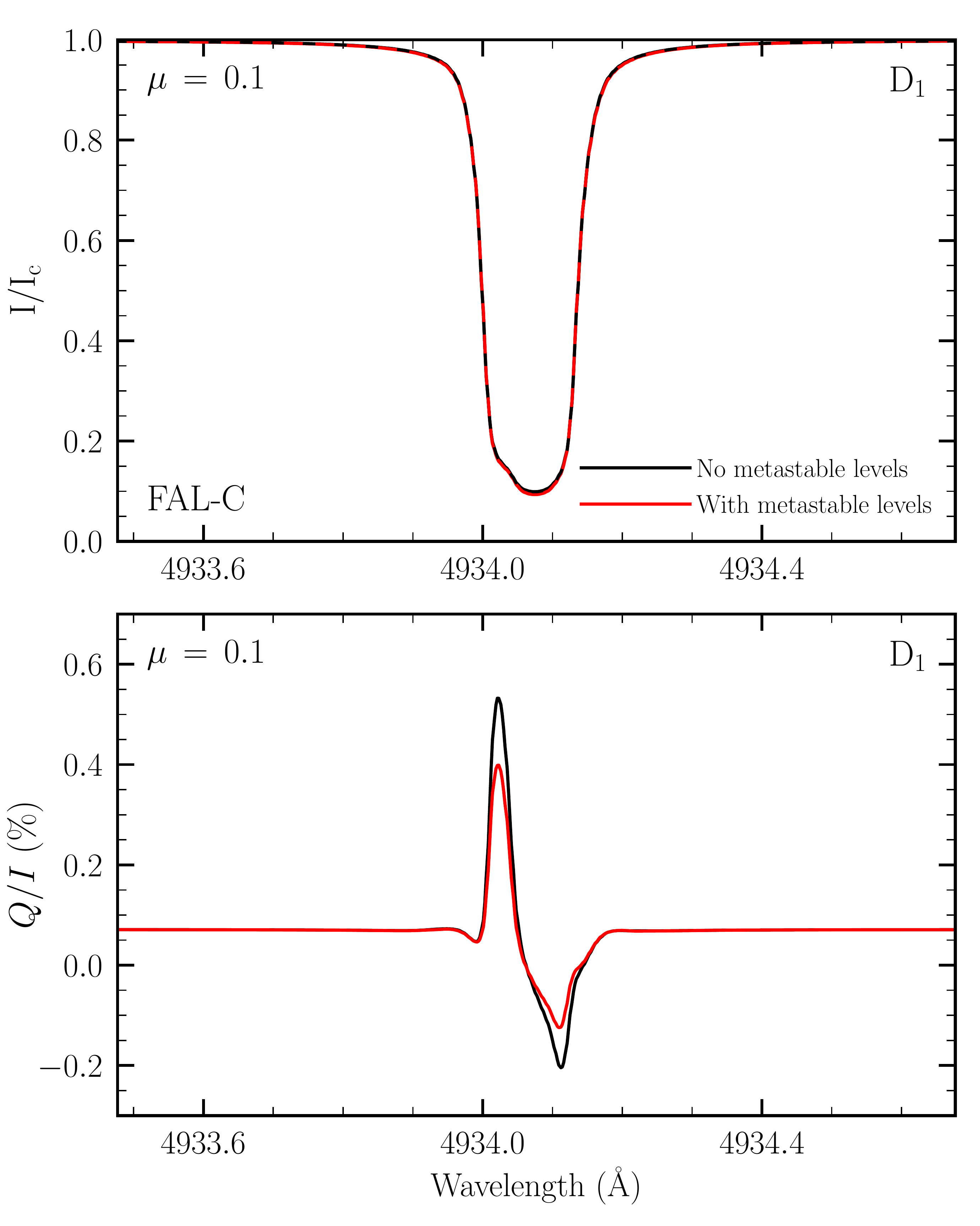} 
\caption{Stokes $I$, normalized to the continuum intensity $I_\mathrm{c}$ (upper panel), and fractional linear polarization 
$Q/I$ profiles (lower panel) of the D$_1$ line as a function of wavelength. The synthetic profiles are obtained from calculations 
using the HanleRT code, accounting for partial frequency redistribution (PRD) effects in the angle-averaged (AA) case and accounting 
for hyperfine structure (HFS) as discussed in the text. 
The black and red curves correspond to calculations including and neglecting the metastable levels, respectively. 
For all the figures presented in the {main text}, the spectral range is $1.2$~\AA\ wide and is centered on the
D$_1$ line. A line of sight (LOS) with $\mu = 0.1$ is taken and the reference direction for positive Stokes $Q$ 
is parallel to the nearest limb.} 
\label{fig::F2-HanleRT-D1-metast}
\end{figure}

In Figure~\ref{fig::F2-HanleRT-D1-metast}, we compare the D$_1$ profiles obtained when including and excluding the $5d\,^2\mathrm{D}$ 
metastable levels in the atomic model. The figure shows the intensity normalized to the continuum intensity at $2$~\AA\ to the red of the line 
center, $I_c$, and the fractional linear polarization pattern $Q/I$. We find an absorption profile in $I/I_c$, which is clearly broadened due to the \gls*{hfs}. 
The inclusion of the metastable levels does not appear to have any impact on the intensity profile ({as long as the fixed lower term population is calculated 
considering the full atomic model}); see also Appendix~\ref{appC}, where the same behavior is found in the corresponding profile for D$_2$. 

The $Q/I$ pattern presents a positive blue peak and a negative red one, whose amplitudes decrease when 
accounting for the metastable levels. This may be attributed to the transfer of population imbalances and quantum interference between magnetic 
sublevels (i.e., atomic polarization) from the $6p\,{}^{2}\mathrm{P}^{\mathrm{o}}_{1/2}$ upper level to the $5d\,^{2}\mathrm{D}_{3/2}$ metastable level. 
Indeed, we note that the $6p\,{}^{2}\mathrm{P}^{\mathrm{o}}_{1/2}$ level only presents atomic polarization due to the spectral structure of the 
incident radiation field. The overall shape of the profile obtained when including or neglecting the metastable levels is very similar, with the blue (red) peak 
remaining positive (negative). This similarity suggests that, if the aim is to qualitatively study the sensitivity of these profiles to specific physical 
mechanisms such as those driven by the magnetic field, one can reasonably model the Stokes profiles of the D$_1$ line with a two-term atomic 
model that neglects the metastable levels (but accounts for the atomic \gls*{hfs}). Indeed, this is the atomic model that is considered in the following 
sections of this work. Regardless, one must be aware that suitably reproducing spectropolarimetric observations of the Ba~{\sc{ii}} 
lines will require the inclusion of such metastable levels. 
We note that the shape of the D$_2$ scattering polarization profile is modified by the metastable levels to a far greater degree than that of D$_1$. 
The discussion of the D$_2$ line can be found in Appendix~\ref{sec-app::FigsD2}. 

\begin{figure}[!h]
  \centering
 \includegraphics[width = 0.485\textwidth]{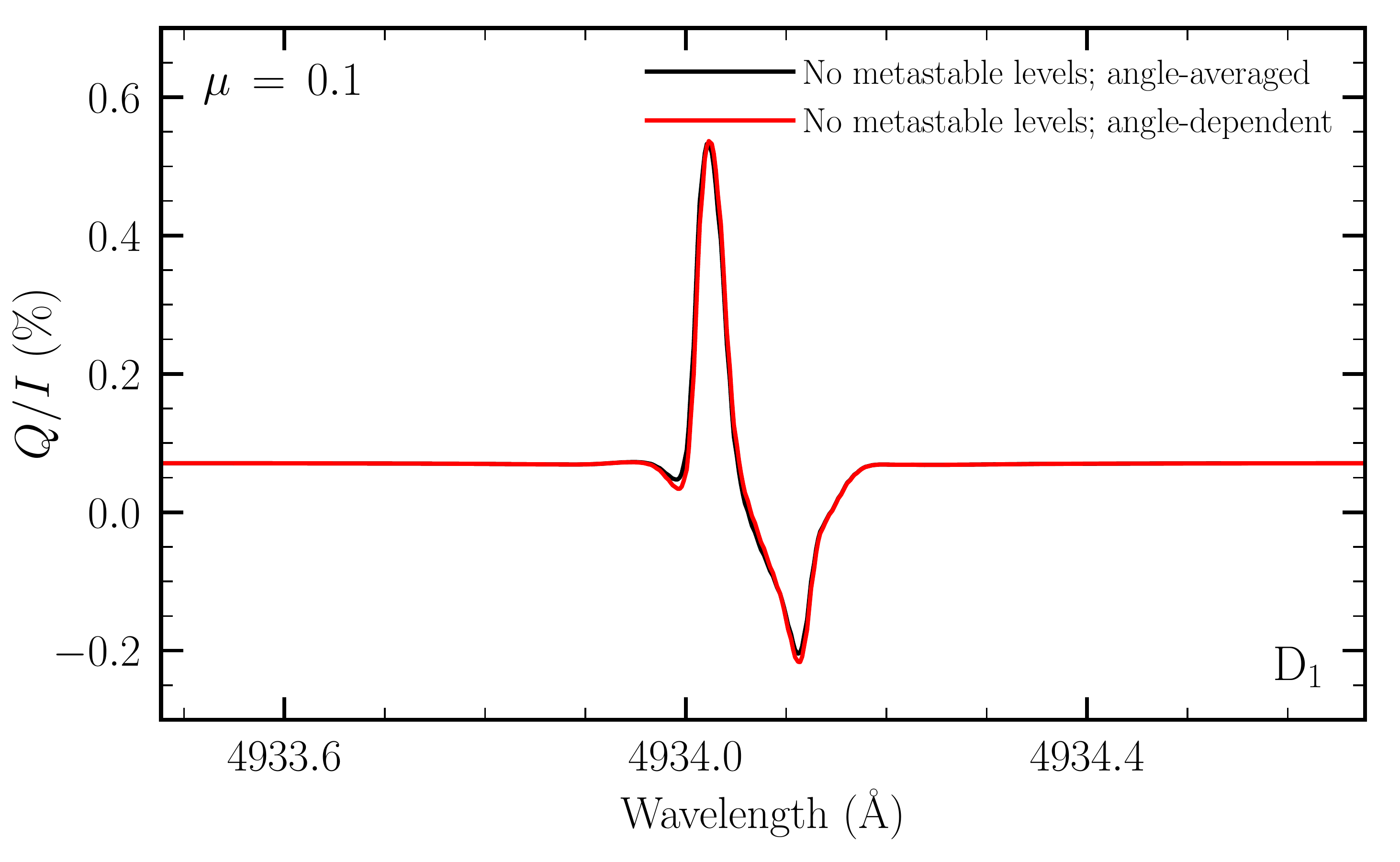}
  \caption{Fractional linear polarization $Q/I$ profiles for the Ba~{\sc{ii}} line as a function of wavelength. The synthetic profiles were 
  computed using HanleRT, accounting for \gls*{prd} effects both under the \gls*{aa} approximation (black curve) and considering the 
  fully angle-dependent case (red curve). In the atomic model, the metastable levels were neglected but the \gls*{hfs} was taken into account.} 
  \label{fig::AAvsAD}
\end{figure}

HanleRT can also solve the non-\gls*{lte} \gls*{rt} problem for polarized radiation accounting for \gls*{prd} effects while relaxing the 
\gls*{aa} approximation (i.e., fully accounting for the frequency-angular coupling due to the Doppler effect). 
Figure~\ref{fig::AAvsAD} shows the comparison between the fractional linear polarization $Q/I$ profiles resulting from calculations 
with and without the \gls*{aa} approximation. For such calculations, we considered a three-level atomic system (i.e., without the 
metastable levels) with \gls*{hfs}. We find a good agreement between the two calculations, which highlights the suitability of the 
\gls*{aa} approximation for modeling the linear polarization pattern of the Ba~{\sc{ii}} D$_1$ line, at least in the absence of magnetic fields. 
This contrasts with the results of the analogous investigation for the Na~{\sc{i}} D$_1$ line reported in 
\cite{Janett+23}, in which such approximation was found to have a clear impact on the shape of the $Q/I$ profile. 
Although it is not shown here, were able to reproduce such findings in the nonmagnetic case using HanleRT.  
Such differences may be attributed to the fact that the \gls*{hfs} splittings of the upper and lower levels of the Ba~{\sc{ii}} 
D$_1$ line (for the isotopes with nonzero nuclear spin) are more than one order of magnitude larger than those of the 
corresponding levels of Na~{\sc{i}}. The separation between the \gls*{hfs} components of the Ba~{\sc{ii}} D$_1$ line is 
proportionally larger, reducing potential spectral overlaps between them due to the Doppler effect. Making the full angle-dependent 
treatment of scattering processes likewise has no impact on the intensity profile of the Ba~{\sc{ii}} D$_1$ line. The synthetic profiles 
presented in the rest of this work were obtained under the \gls*{aa} approximation. 

\section{The impact of the HFS, atmospheric model, and magnetic fields} 
\label{sec::numerical} 
In this section we continue investigating the formation of the intensity and polarization profiles of the Ba~{\sc{ii}} D$_1$ line in 
optically thick atmospheres. The numerical approach considered here assumes a two-term atomic model and does not allow 
for the inclusion of the metastable levels, whose impact on the amplitude of the D$_1$ scattering polarization is not negligible. 
On the other hand, it does allow for investigations accounting for the quantum interference between states within the same 
\gls*{fs} and/or \gls*{hfs} levels while in the presence of magnetic fields of arbitrary strength and orientation.  
\subsection{Numerical approach}
\label{sec-sub::form-num-framework}
The synthetic Stokes profiles presented in this section were obtained through the following two-step approach. 

\noindent {\textit{Step 1}}: 
Compute a number of quantities to be used as input for the second step, including the collisional rates, continuum quantities, and the 
population of the ground term $\mathcal{N}_\ell$. Such calculations are carried out by solving the 
non-\gls*{lte} problem without polarization using the RH code of \cite{Uitenbroek01}. The considered atomic system includes the 
ground level of Ba~{\sc{iii}} and the five levels of Ba~{\sc{ii}} discussed in Section~\ref{sec::problem}, but not the \gls*{hfs}. 
Because the metastable levels are included in the calculations in this step, they yield a more accurate value for
$\mathcal{N}_\ell$ than when considering a two-term atomic system. More details on such calculations can be found in Appendix~\ref{sec-app::AtomicQuantTables}. 

\noindent {\textit{Step 2}}: 
Obtain the synthetic Stokes profiles for the D lines by solving the non-\gls*{lte} \gls*{rt} problem in the polarized case via the numerical code 
described in \cite{AlsinaBallester+22}. It is suitable for a two-term atomic system and thus does not account for the metastable levels of Ba~{\sc{ii}}, 
but it can include the \gls*{hfs} of the odd isotopes. Unless otherwise noted, all seven stable isotopes are considered.\footnote{Each coefficient of the 
\gls*{rt} equation can be taken as a linear combination of the contribution from each single isotope, weighted by its relative abundance 
\citep[e.g.,][]{AlsinaBallester22}.} The code can account for scattering polarization with both \gls*{prd} effects under the \gls*{aa} approximation 
and magnetic fields in the incomplete Paschen-Back effect regime. 
The ground term is assumed not to have atomic polarization, because elastic collisions with neutral hydrogen are expected to suppress the ground 
level atomic polarization, as they do for the metastable levels 
\citep[see][]{Derouich08}. 
Thus, each of the \gls*{hfs} levels of the ground term is populated according to the total $\mathcal{N}_\ell$ and its corresponding statistical weight. $\mathcal{N}_\ell$ is 
kept fixed throughout the iterative \gls*{rt} calculation for this step and, because all the \gls*{rt} coefficients are proportional to this value, the problem is
linear.\footnote{For the problem to be linear, stimulated emission must also be neglected, which is a very good approximation for the wavelengths of interest. 
Indeed, this assumption is made in the derivations of the \gls*{prd} formalisms on which HanleRT and the code described in \cite{AlsinaBallester+22} are based.} 
The thermal line emissivity is computed as explained in \cite{AlsinaBallester+22}. The potential impact of the collisional transfer of atomic polarization 
between different \gls*{fs} or \gls*{hfs} levels is beyond the scope of this work, and was not taken into account in the calculations for the profiles presented below.  
\begin{figure}[!h]
\centering 
\includegraphics[width = 0.485\textwidth]{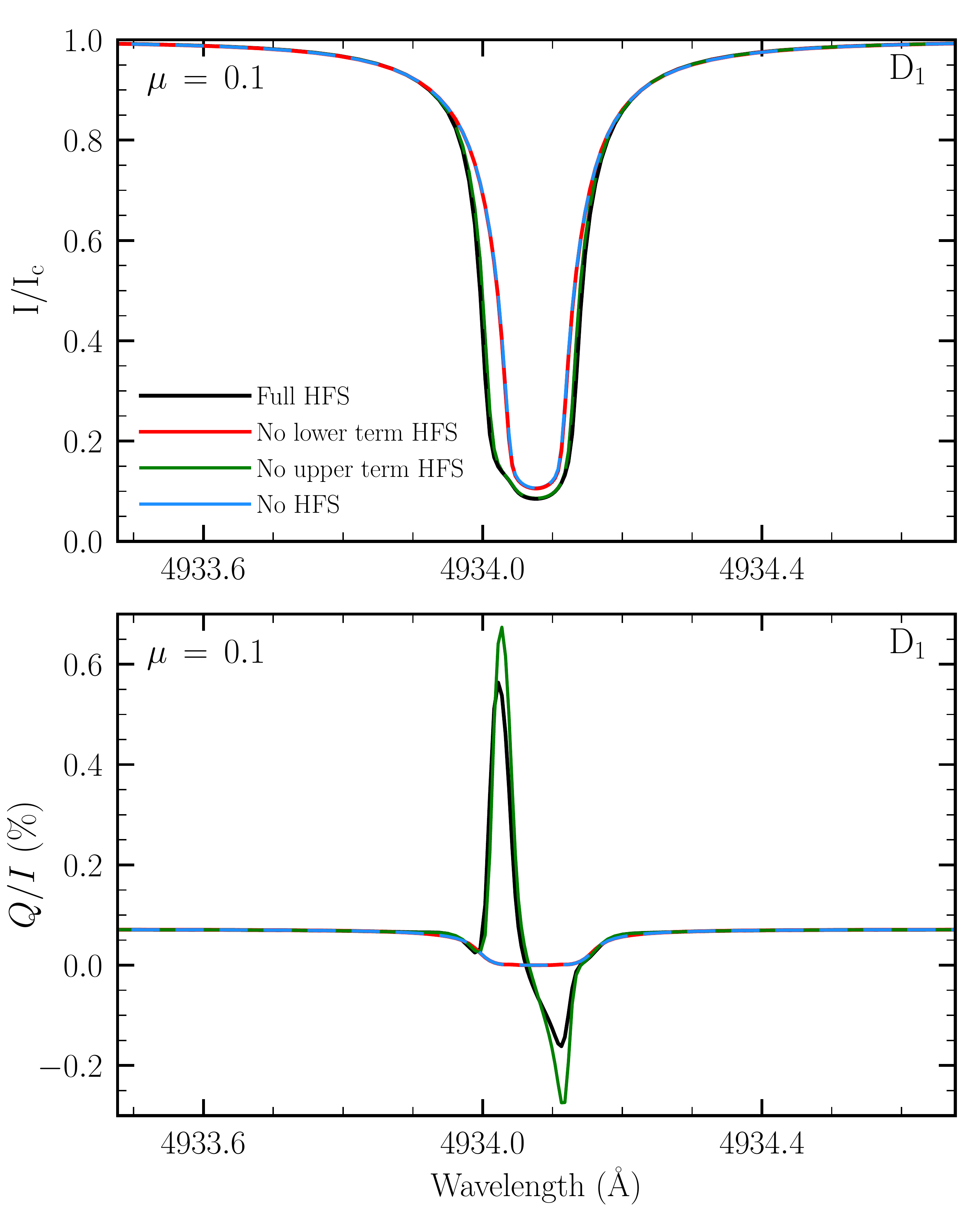}
\caption{Stokes $I$, normalized to the continuum intensity $I_\mathrm{c}$ (upper panel), and $Q/I$ (lower panel) profiles of the D$_1$ line as a 
function of wavelength. In all the figures presented in this section, the profiles were calculated following the approach described at the beginning of this section. 
The colored curves indicate the results of calculations with different treatments of the hyperfine structure (HFS): fully accounting for it (black), accounting 
only for that of the upper (red) or lower (green) term, or neglecting it entirely (blue). Overlapping curves are dashed for the sake of visibility.} 
\label{fig::F3-HFS}%
\end{figure} 

Because we are considering \gls*{1d} atmospheric models without bulk velocities, the problem is axially symmetric along the local 
vertical (except in the presence of inclined magnetic fields; see Section~\ref{sec-subsub::horiz}). Under such symmetry conditions, and taking the 
reference direction for positive Stokes $Q$ parallel to  the nearest limb, no Stokes $U$ or $V$ are produced and thus the corresponding figures are not shown. 

\subsection{The impact of HFS} 
\label{sec-sub::num-hfs} 
The {black curves in Figure~\ref{fig::F3-HFS} represent the D$_1$ intensity profile normalized to $I_c$ (top panel) and the 
$Q/I$ profile (bottom panel), obtained as described above. Such calculations were carried out in the absence of magnetic 
fields, considering the FAL-C model and accounting for the \gls*{hfs} of the odd isotopes, as well as for the quantum 
interference between all states of the upper term}. 
Like in the case of the $Q/I$ profile obtained with HanleRT when neglecting the metastable levels 
(see Section~\ref{sec-sub::metastable}), we find a positive $Q/I$ blue peak and a negative red one. The amplitude 
of the blue peak is roughly $0.6\%$ (slightly larger than that found with HanleRT) and the amplitude of the red one is 
just above $0.15\%$ (slightly smaller). 

Figure~\ref{fig::F3-HFS} also highlights the impact of the \gls*{hfs} of barium on the intensity and scattering polarization 
patterns of the D$_1$ line, by presenting a comparison between the profiles discussed in the previous paragraph, in which 
the \gls*{hfs} splitting was fully taken into account, and the profiles obtained by neglecting it in the 
$6s\,^{2}\mathrm{S}$ ground term (red curve), the $6p\,^{2}\mathrm{P}^{\mathrm{o}}$ upper term (blue curve), or 
both (green curve). Such splittings were neglected by setting to zero the corresponding $\mathcal{A}$ and $\mathcal{B}$ \gls*{hfs} coefficients
(see Appendix~\ref{sec-app::AtomicQuantTables}). 

Only $18\%$ of barium atoms have \gls*{hfs}, but this already leads to a substantial broadening of the D${}_1$ intensity 
profile \citep[in agreement with][]{BelluzziTrujilloBueno13}. This broadening is mostly due to the \gls*{hfs} splitting of the ground term, 
which is more than one order of magnitude larger than that for the upper level of D${}_1$. The two-peak scattering polarization pattern 
can only be reproduced by accounting for the \gls*{hfs} and, specifically, that of the ground term. If the latter splitting is neglected, the 
spectral window spanned by the various \gls*{hfs} components of the D$_1$ line is very small and the radiation field is effectively flat within this range. 
As a result, the key mechanism pointed out by \cite{BelluzziTrujilloBueno13}, through which scattering polarization is produced in this 
intrinsically unpolarizable line, is inhibited. On the other hand, accounting for the \gls*{hfs} of the ground term but neglecting that of 
the upper level of D$_1$ leads to an enhancement of the amplitude of the polarization peaks. This enhancement occurs because the quantum interference 
between the various $F$ levels is maximum if there is no energy separation between them. 

We also carried out calculations in which we fully accounted for the \gls*{hfs} but neglected the quantum interference between states pertaining 
to different $J$ levels of the upper term (i.e., $J$-state interference) and to different $F$ levels of the same $J$ level of the upper term (i.e., $F$-state 
interference), following Appendix~C.7 of \cite{AlsinaBallester+22}. Neither $J$- nor $F$-state interference have an appreciable impact on the 
scattering polarization of the D$_1$ line and the corresponding profiles are thus not shown. Such results were expected, because the 
separation between the upper \gls*{fs} levels of the D$_1$ and D$_2$ lines is extremely large and even the \gls*{hfs} splitting in the upper level of 
D$_1$ is considerably more than one order of magnitude larger than the natural width of the line. 

\subsection{The sensitivity to the atmospheric model} 
The semiempirical \gls*{1d} atmospheric models considered in this work are representative of spatial averages 
of specific regions of the solar atmosphere and thus cannot account for the full \gls*{3d} complexity of the 
real solar atmosphere. Moreover, bulk velocities are not taken into account in this work, despite the dynamic nature
of the Sun. Nevertheless, our modeling can provide valuable insights into the physics that shape the intensity 
and polarization patterns of non-\gls*{lte} lines 
\citep[see, e.g., ][in which the Stokes profiles of the Ba~{\sc{ii}} D$_2$ line were synthesized considering various 
semiempirical models and compared them with observations]{Faurobert+09,Smitha+13}. 
Here, we present the synthetic intensity and $Q/I$ profiles of the Ba~{\sc{ii}} D$_1$ line obtained with several FAL models other 
than FAL-C, which was used in the calculations presented above and is representative of an average region of the quiet solar atmosphere. 
The other considered semiempirical models are FAL-A, which represents relatively faint internetwork regions of 
the quiet Sun; FAL-F, representative of particularly bright network regions of the quiet Sun; and FAL-P, which corresponds to a typical plage region. 
We also considered FAL-X, which is representative of an average region of the quiet solar atmosphere, but with considerably lower temperatures 
than FAL-C in the photosphere and up to the middle chromosphere. 
Throughout the entire wavelength range taken for the problem (which includes the D$_1$ and D$_2$ lines and their nearby continuum), the intensity 
is highest for model P, then F, then C and X (having very similar values for both models), and is lowest for model A. 
However, the D$_1$ intensity profiles, when normalized to $I_c$, present a remarkably similar shape for all considered models. The most 
appreciable differences concern the width of the wings, but even these are minor and are thus not shown here. 

\begin{figure}[!h]
 \centering
 \includegraphics[width = 0.485\textwidth]{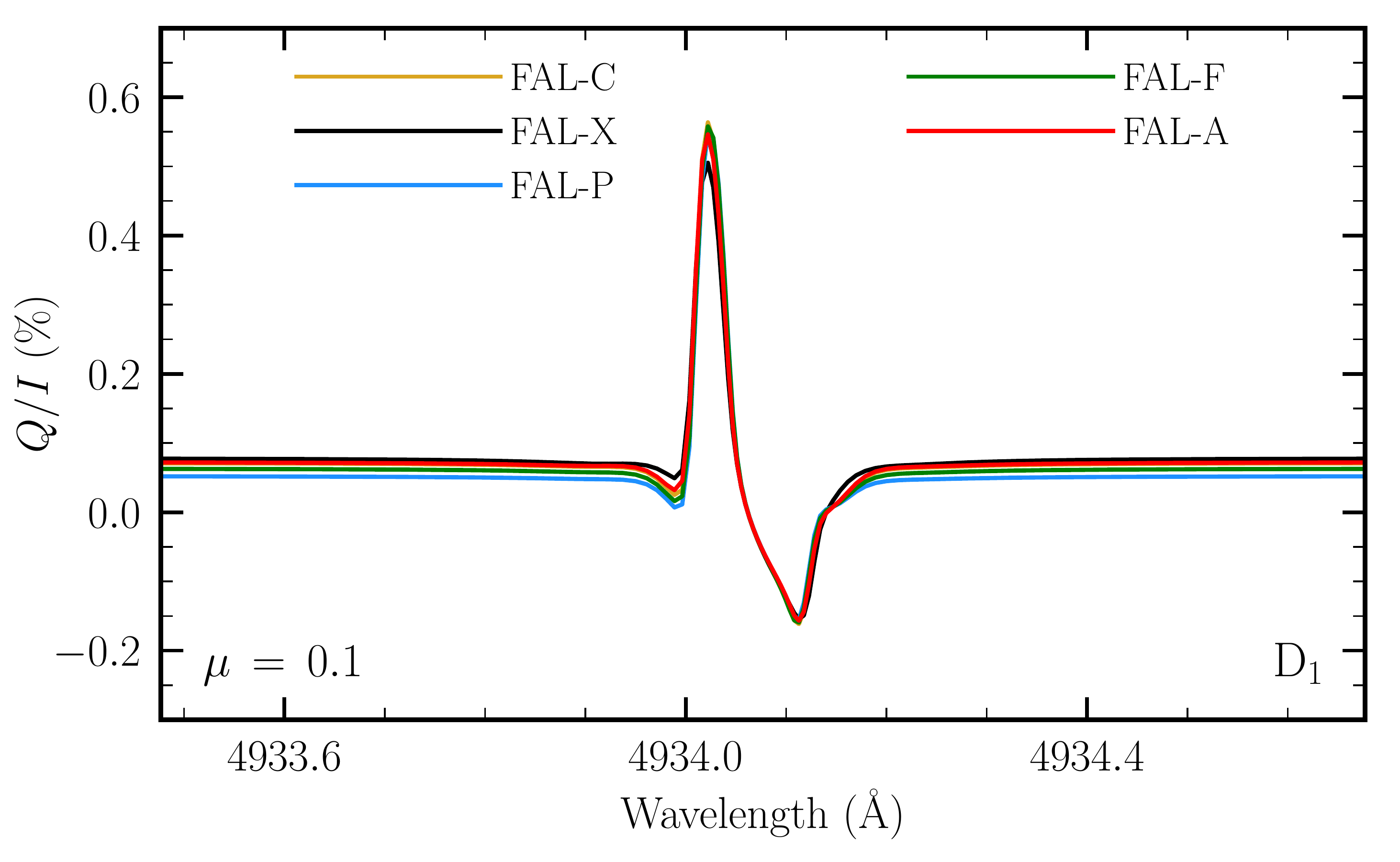}
  \caption{Synthetic $Q/I$ profiles of the D$_1$ line as a function of wavelength, obtained considering different \gls*{1d} 
  emiempirical atmospheric models. The colored curves correspond to calculations considering the atmospheric models 
  indicated in the legend.} 
  \label{fig::FALmodel}%
\end{figure} 
The differences between the D$_1$ fractional scattering polarization profiles for the various considered models are also 
quite modest, as can be seen in Figure~\ref{fig::FALmodel}. Indeed, the largest differences are found between the $Q/I$ profile obtained 
considering the FAL-X model (with a maximum amplitude of $\sim\!0.50 \%$ in the blue peak) and the other models 
(whose blue peaks reach amplitudes between $\sim\!0.54$ and $0.56 \%$). 

In order to replicate the spectral smearing due to large-scale velocities typical of the lower chromosphere and the finite 
resolution of a typical instrument, we convolved the synthetic profiles with a Gaussian function with a FWHM of $70$~m\AA . 
After such smearing, the $Q/I$ profiles for the various models are indistinguishable at the plot level (figure not shown), and 
differences are only appreciable in the continuum polarization. 
We emphasize that even the smeared D$_1$ profiles still present a positive blue peak and a negative red one, in contrast to 
the observations reported in {Figure 3 of \cite{Stenflo+2000}. None of the physical ingredients considered in this paper (including 
metastable levels, angle-dependent \gls*{prd}, and the magnetic fields, discussed below) can produce a positive red $Q/I$ peak 
in the D$_1$ line. For further progress in this respect, we need high-precision spectropolarimetric observations of this line and 
to include in our theoretical modeling the non-coherent continuum scattering investigated by \cite{delPinoAleman+14a,delPinoAleman+14b}.} 

Our results indicate that the D${}_1$ line is largely insensitive to the thermodynamical structure of the solar atmosphere and, 
thus, observable variations in its scattering polarization should be attributed to other factors, such as the presence of a magnetic field. 
In the future, it will be of interest to investigate the possible sensitivity of the D$_1$ intensity and polarization signals to 
changes in atmospheric models that are \gls*{3d} rather than \gls*{1d} and dynamic rather than static. 

\subsection{Magnetic fields}
\label{sec-sub::magnetic} 
{The presence of a magnetic field modifies the energy of the magnetic states $f$ of the Ba~{\sc{ii}} atom (as illustrated in 
Figure~2 of \citealt{Belluzzi+07} for the $^{137}$Ba isotope; see also Figure~2 of \citealt{Belluzzi+07b}). This impacts the polarization 
of the spectral lines by producing a shift in the $\pi$ and $\sigma$ components of the line\footnote{Throughout this work we refer to {such
spectral line polarization as} due to the Zeeman effect, even when the shifts {in the $\sigma$ and $\pi$ components 
do not depend linearly} on the magnetic field because of mixing between states with different $J$ or $F$.} 
and by modifying the quantum interference between $f$ states}. 
{A main point of interest in this work is to evaluate the magnetic sensitivity of the polarization patterns of the Ba~{\sc{ii}} D$_1$ line, thus 
providing valuable insights into the potential of this spectral line for diagnostics of chromospheric magnetic fields.} 
Here we present a series of numerical experiments, considering magnetic fields of increasing strength that are 
either isotropically distributed (see Sect.~\ref{sec-subsub::isotropic}) or deterministic (see Sect.~\ref{sec-subsub::horiz}). 

\subsubsection{Tangled magnetic fields}
\label{sec-subsub::isotropic}
We analyze the sensitivity of the D$_1$ scattering polarization to magnetic fields whose orientation changes at scales smaller than the 
mean free path of the photons of the line \citep[following Appendix C.6 of][{where such fields are called micro-structured}]{AlsinaBallester+22}, 
with no preferred direction. In particular, we consider such fields with an isotropic distribution of orientations and a fixed strength, which we 
hereafter refer to as tangled magnetic fields. 
Such fields do not break the axial symmetry of the problem, and thus they do not give rise to any Stokes $U$ or $V$ signal. 
\begin{figure}[!h]
  \centering
 \includegraphics[width = 0.485\textwidth]{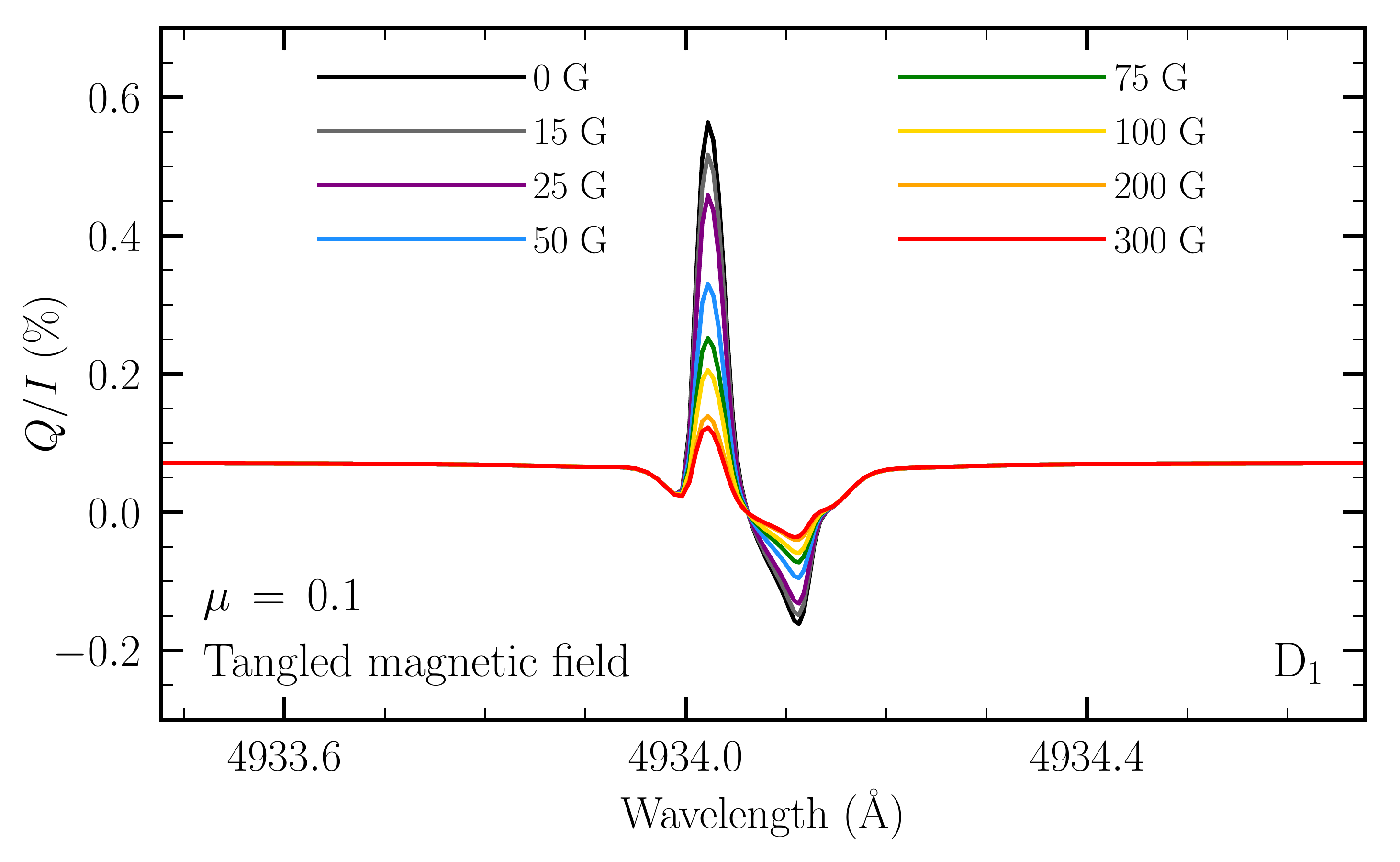}
  \caption{Synthetic $Q/I$ profiles of the D$_1$ line as a function of wavelength obtained in the presence of tangled magnetic fields. 
  The various colored curves represent the calculations carried out considering the field strengths indicated in the legend.} 
  \label{fig::D1Microturb} 
\end{figure} 
{We carried out calculations for tangled magnetic fields up to $500$~G, although in Figure~\ref{fig::D1Microturb} we only show 
the $Q/I$ profiles obtained for field strengths up to $300$~G.} The intensity profile does not change appreciably within 
the considered range {of field strengths}, for which the magnetic splitting is much smaller than the Doppler width, and thus 
the corresponding figure is not shown. The amplitude of the D$_1$ scattering polarization begins to decrease appreciably in 
the presence of fields with strengths of about $15$~G. As the field strength increases further, the polarization amplitude 
decreases monotonically, but this trend begins to halt at about $200$~G. 
Although it is not shown in the figure, we also verified that further increases in the field strength beyond $300$~G barely 
modify the linear polarization amplitude (i.e., saturation is reached). At saturation, the $Q/I$ amplitude of the red peak 
is $\sim\!0.12\%$, which is {approximately} $1/5$ of the one obtained in the absence of magnetic fields (roughly $0.60\%$), 
as expected for the saturation value for isotropic micro-structured fields in a two-level atom \citep[e.g.,][]{TrujilloBuenoMansoSainz99}. 
Recalling that the D$_1$ scattering polarization pattern is produced only by the $\sim\!18\%$ of barium isotopes that have 
nonzero nuclear spin ($^{135}$Ba and $^{137}$Ba), we focus the discussion on such isotopes and their \gls*{hfs}. 
We verified numerically that neglecting the magnetic splitting of the ground level does not change the D$_1$ 
scattering polarization; its magnetic sensitivity can be mainly attributed to the splitting of the upper level. 
Indeed, the Larmor frequency at $15$~G is close to $1/5$ of the natural width of the line's upper level; at 
that point the splitting between the magnetic states $f$ of any given \gls*{hfs} level of 
$6p\, ^2\mathrm{P}^\mathrm{o}_{1/2}$ {becomes} large enough that their interference appreciably decreases, 
reducing their scattering polarization {(the Hanle effect)}. 
As the magnetic field increases, so does the splitting between the $f$ states and thus the interference between them 
becomes weaker. At saturation field strengths, the separation between $f$ states of the same \gls*{hfs} {level} 
is large enough that the interference between them is negligible. 

\subsubsection{Deterministic magnetic fields}
\label{sec-subsub::horiz}
We also investigate the case of deterministic magnetic fields (i.e., those with a fixed direction rather 
than an isotropic distribution of orientations), considering {first} the specific case of horizontal 
magnetic fields contained in the plane defined by the local vertical and the \gls*{los}. 
For an \gls*{los} with $\mu = 0.1$, such fields {are almost longitudinal}. {In this case}, the problem is 
no longer axially symmetric and nonzero $U$ and $V$ signals can arise. 
\begin{figure}[!h]
 \centering
\includegraphics[width = 0.485\textwidth]{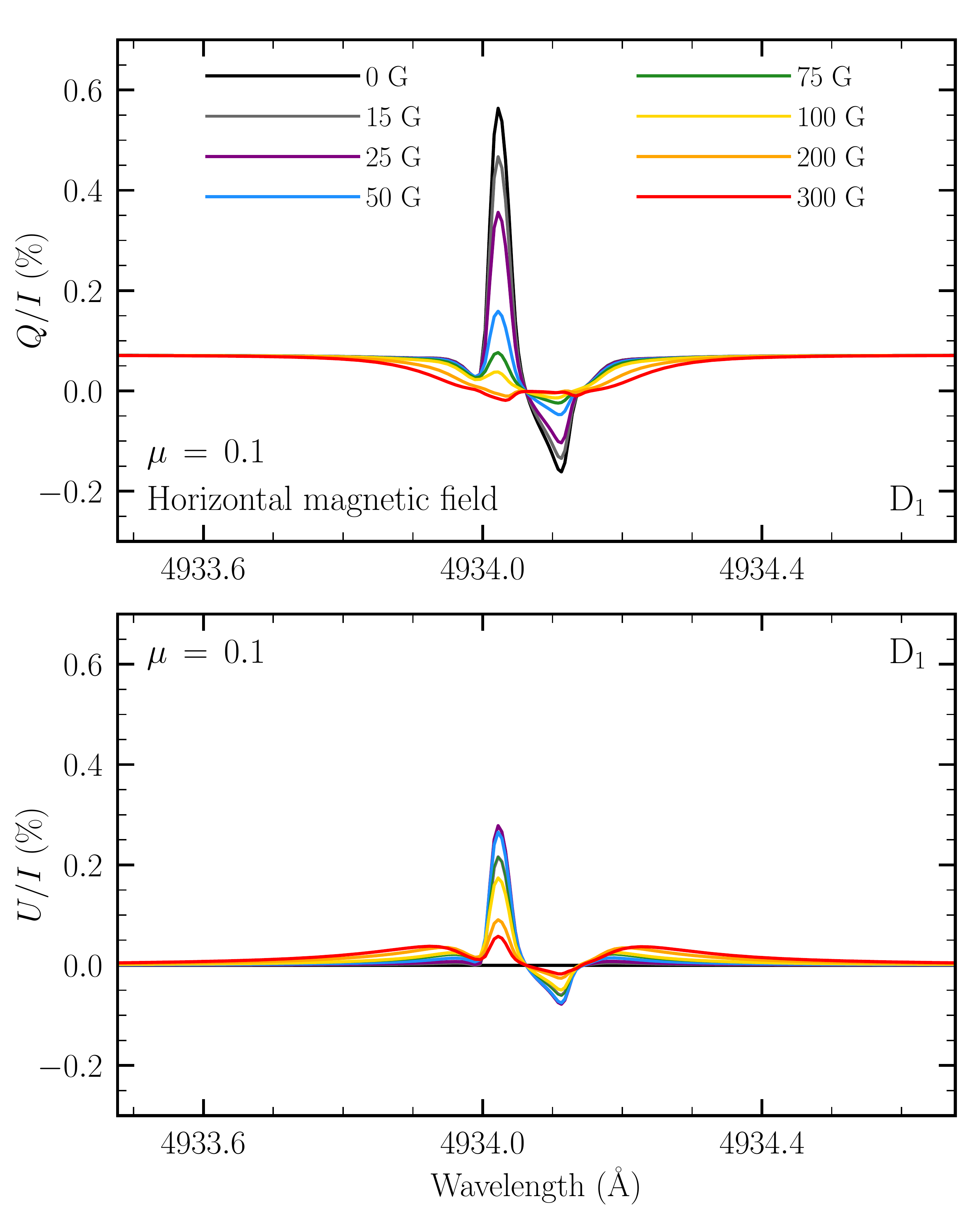}
 \caption{Synthetic $Q/I$ {(upper panel)} and $U/I$ {(lower panel)} profiles of the D$_1$ line as a function of wavelength, 
  obtained in the presence of deterministic horizontal magnetic fields contained within the plane given by the local vertical and 
  the \gls*{los}. The various colored curves represent the calculations carried out in the presence of fields with the strengths 
  indicated in the legend.} 
  \label{fig::D1-QU-Horiz}%
\end{figure} 
Figure~\ref{fig::D1-QU-Horiz} shows a series of $Q/I$ and $U/I$ profiles obtained in the presence of horizontal magnetic 
fields with a positive projection onto the \gls*{los} and the same strengths considered in Section~\ref{sec-subsub::isotropic}. 
Such magnetic fields reduce the amplitude of the $Q/I$ signal to a greater degree than tangled fields of the same strengths. 
For magnetic fields close to saturation (of $200$~G or stronger), a depolarization pattern is found in $Q/I$. For this geometry, 
the Hanle effect also gives rise to a $U/I$ signal, whose amplitude increases with magnetic field strength until about $35$~G. 
For stronger fields, the $U/I$ amplitude instead decreases as the magnetic field reduces the interference between $f$ states. 
 
\begin{figure}[!h]
  \centering
  \includegraphics[width = 0.485\textwidth]{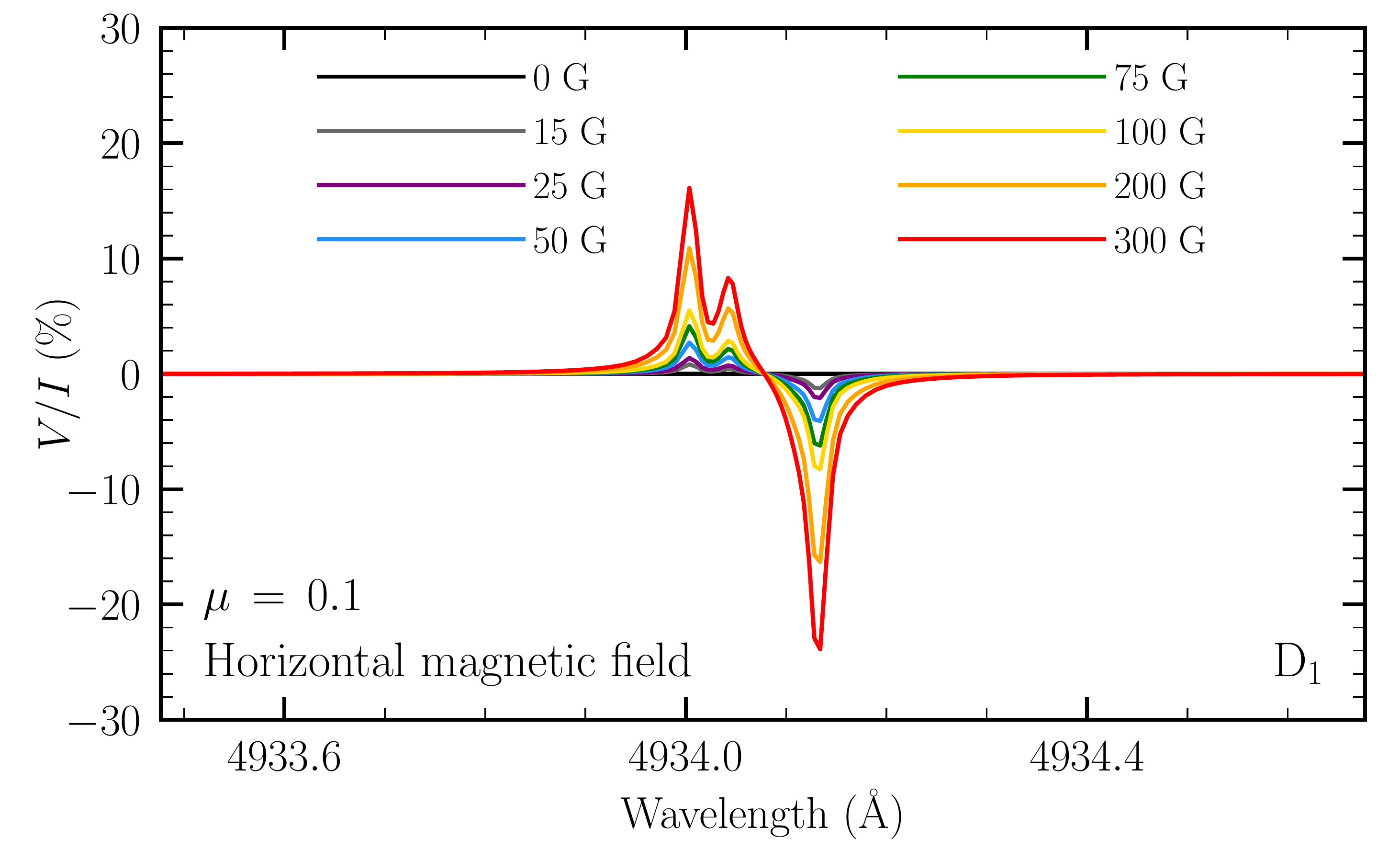}
  \caption{Synthetic $V/I$ profiles of the D$_1$ line as a function of wavelength, obtained in the presence of {the same} 
  horizontal magnetic fields {considered} in the previous figure.} 
  \label{fig::D1-V-Horiz}%
\end{figure} 
 
The considered horizontal fields {also} give rise to a $V/I$ pattern with two positive peaks to the blue of the line center 
and a negative peak to the red, as illustrated in Figure~\ref{fig::D1-V-Horiz}. In the presence of a 15~G magnetic field, 
the amplitudes of these peaks reach roughly $1\%$ and they increase linearly with field strength within the range 
considered in this work. This is the behavior typically associated with the Zeeman effect. 
The double-peak feature found to the blue of the line center is due to the large \gls*{hfs} splitting of the upper and 
lower levels of the D$_1$ line. Indeed, we verified that a $V/I$ pattern with a single blue peak is produced instead 
when the \gls*{hfs} is neglected. This contrasts with the $V/I$ signals found for the K~{\sc{i}} D$_1$ line, which arises from a 
transition between levels with the same $J$ and $F$ quantum numbers, but which could be suitably modeled without 
accounting for the \gls*{hfs} \citep[see][]{AlsinaBallester22} because its splitting is much smaller than for the analogous 
levels of Ba~{\sc{ii}}. We also verified that one cannot suitably apply the magnetograph formula 
\citep[e.g., Section~9.6 of][]{BLandiLandolfi04} to the Ba~{\sc{ii}} D$_1$ line, calculating the Land\'e factors according 
to $L$-$S$ coupling (neglecting \gls*{hfs}). 

Finally, we evaluated the suitability of the so-called linear Zeeman approximation, that is, neglecting the off-diagonal 
elements of the magnetic Hamiltonian, which are responsible for the mixing between states with different $J$ or $F$ eigenstates. 
We verified numerically that, although the mixing between $J$ states can be safely neglected, neglecting the mixing between $F$ 
states substantially underestimates the amplitude of the circular polarization patterns. The unsuitability of the linear Zeeman approximation had 
also been reported for spectral lines such as H~{\sc{i}} Lyman-$\alpha$ \citep[see][Appendix A]{AlsinaBallester+19}, the Mn~{\sc{i}} 
resonance multiplet around $2800$~\AA\ \citep{delPinoAleman+22}, or the K~{\sc{i}} D lines \citep{AlsinaBallester22}, 
for which the $J$ or $F$ mixings due to the \gls*{ipb} effect are significant. 

\begin{figure}[!h]
 \centering
 \includegraphics[width = 0.485\textwidth]{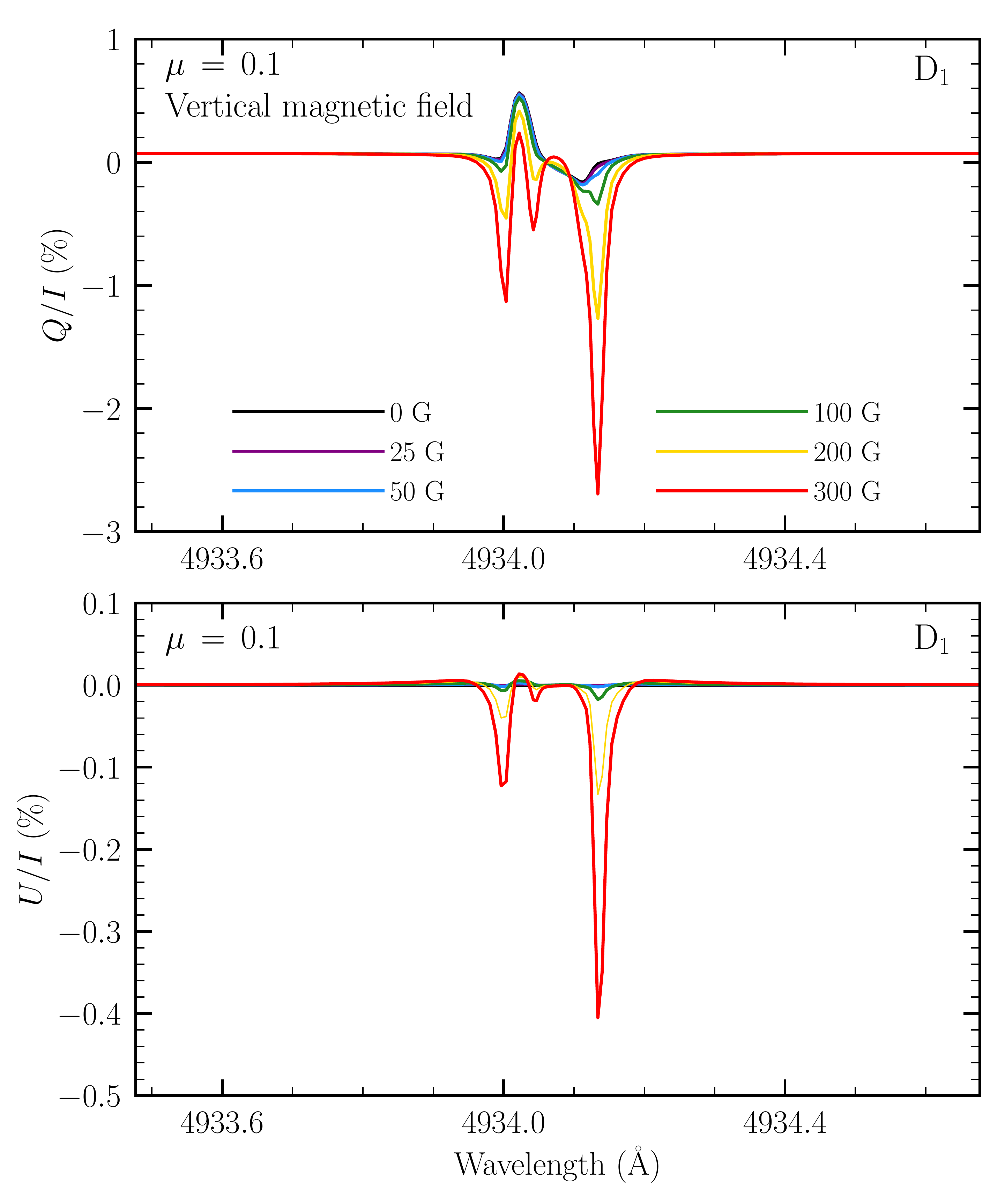}
 \caption{Synthetic $Q/I$ {(upper panel)} and $U/I$ {(lower panel)} profiles of the D$_1$ line as a function of wavelength, obtained 
 in the presence of deterministic vertical magnetic fields. The various colored curves represent the calculations carried in the presence 
 of fields with the strengths as indicated in the legend.}
  \label{fig::D1-QU-Vert}%
\end{figure} 
We also considered the case of a vertical deterministic magnetic field, which begins to appreciably impact the linear polarization 
patterns in the presence of magnetic fields of about $200$~G, as can be seen in Figure~\ref{fig::D1-QU-Vert}. Unlike the aforementioned 
horizontal fields, such vertical fields do not impact the interference between the $f$ states that are degenerate in the nonmagnetic case 
(the Hanle effect does not operate). Moreover, the energies of the two upper \gls*{hfs} levels of the D$_1$ line are too far apart for the 
interference between them to play a meaningful role. Instead, the linear polarization signals can be attributed to the Zeeman effect due 
to the transverse component of the magnetic field. 
\begin{figure}[!h]
  \centering
    \includegraphics[width = 0.485\textwidth]{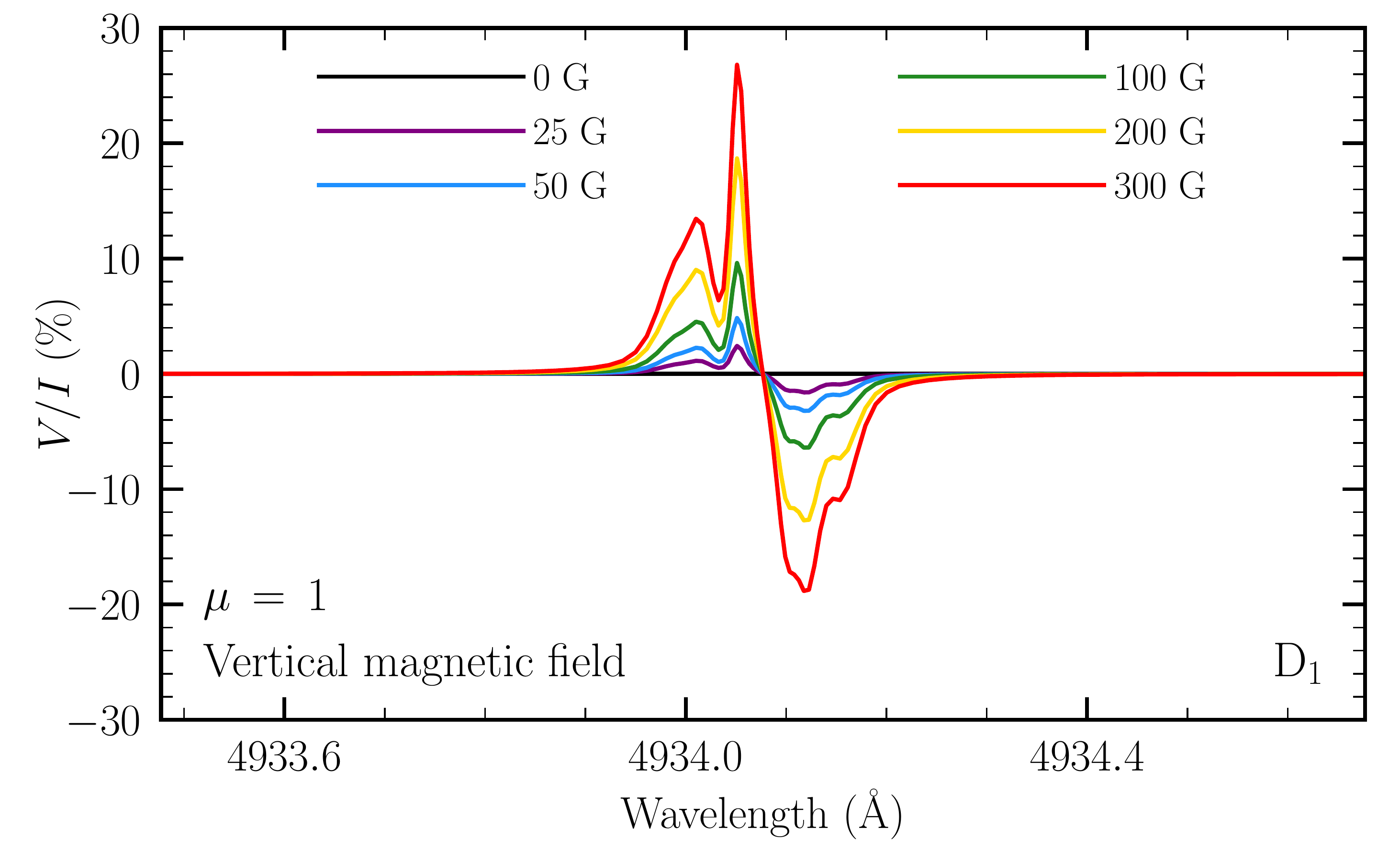}
  \caption{Synthetic $V/I$ profiles of the D$_1$ line as a function of wavelength, obtained in the presence of the same vertical magnetic 
   fields considered in the previous figure. In order to consider a larger longitudinal component of the magnetic field, for this figure we 
   took an \gls*{los} with $\mu = 1$.}   
  \label{fig::D1-V-Vert}%
\end{figure} 
When considering \gls*{los}s close to the disk center, vertical magnetic fields are close to longitudinal and thus give rise to a $V/I$ 
pattern whose amplitude is proportional to the longitudinal component, as shown in Figure~\ref{fig::D1-V-Vert}. Although the shape 
of the $V/I$ profile is noticeably different from the one shown in Figure~\ref{fig::D1-V-Horiz} -- presenting far wider wing lobes, for 
instance -- there is also a clear double-peak feature due to the \gls*{hfs}. Of course, the magnetograph formula and the linear Zeeman 
approximation are not suitable for this geometry, either. 
 
\section{Conclusions} 
\label{sec::conclusions} 
In this work, we carried out a series of numerical experiments to identify the main physical mechanisms that shape  
the intensity and polarization patterns of the D$_1$ line. We obtained the Stokes profiles of these lines through 
non-\gls*{lte} \gls*{rt} calculations, considering semiempirical \gls*{1d} models of the solar atmosphere and atomic 
models that account for both the D$_1$ and D$_2$ lines. Our modeling included both \gls*{prd} effects and the 
\gls*{hfs} of the barium isotopes with nuclear spin ($18$~\% of the total by abundance). This allowed us to study the 
scattering polarization arising from the spectral structure of the anisotropic radiation field over the wavelength interval 
spanned by the various \gls*{hfs} components of D$_1$. In order to consider relatively large scattering polarization signals, 
we displayed the resulting Stokes profiles at an \gls*{los} with $\mu = 0.1$. 

{Here we} evaluated the impact of the metastable levels on the D$_1$ line {in the nonmagnetic case} using the HanleRT 
non-\gls*{lte} code \citep{delPinoAleman+16,delPinoAleman+20}. This \gls*{rt} code is designed for a multi-term 
atomic system without \gls*{hfs}, but through some formal substitutions it {can} incorporate a multi-level atom 
with \gls*{hfs}, which neglects the quantum interference between \gls*{fs} levels but is otherwise suitable in the absence 
of magnetic fields. Although the inclusion of the metastable levels appreciably decreases the amplitude of the $Q/I$ 
pattern of D$_1$, they have little impact on its shape. We also verified with HanleRT that the D$_1$ scattering polarization 
signal can be suitably modeled making the \gls*{aa} approximation. 

For the rest of our investigation, we considered synthetic Stokes profiles calculated {using} the non-\gls*{lte} \gls*{rt} 
code described in \cite{AlsinaBallester+22}. Although this code can account for \gls*{prd} effects only under the 
\gls*{aa} approximation and cannot account for the metastable levels, it can jointly include the \gls*{hfs} of the odd 
isotopes of Ba~{\sc{ii}} and magnetic fields of arbitrary strength and orientation. We find that the very large \gls*{hfs} of the ground  
term substantially broadens the D$_1$ intensity profile and is responsible for its two-peak scattering polarization pattern. 
We also verified that the quantum interference between \gls*{fs} or \gls*{hfs} levels has no significant impact on the linear polarization. 

Interestingly, our numerical experiments considering the various FAL semiempirical atmospheric models (which present different 
stratifications for parameters such as temperature or density) reveal that neither the D$_1$ continuum-normalized intensity nor the 
$Q/I$ fractional linear polarization pattern are strongly sensitive to the different models. Thus, most of the variation of its linear 
polarization can instead be attributed to the magnetic field, enhancing its value for diagnostics of chromospheric magnetic fields. 
In the future, the variation of the intensity and polarization of the D$_1$ line should be investigated when considering dynamic 
models of the solar atmosphere that fully account for its \gls*{3d} complexity. 

In this work we considered tangled and deterministic magnetic fields with strengths of up to $500$~G, although we did not
show the Stokes profiles obtained for fields stronger than $300$~G. The considered fields have no appreciable impact on the
intensity profile. On the other hand, the linear scattering polarization is clearly sensitive to tangled or horizontal magnetic fields 
of roughly $15$~G or stronger via the Hanle effect. Tangled magnetic fields of increasing strength progressively depolarize 
the $Q/I$ signal until reaching saturation at about $300$~G; the $Q/I$ saturation amplitude is approximately $1/5$ of the 
amplitude in the nonmagnetic case. Deterministic horizontal magnetic fields have a stronger depolarizing effect than tangled 
fields of the same strength and, at saturation, present an almost completely depolarized signal. 
In the presence of such fields, the problem is no longer axially symmetric, and thus they give rise to a $U/I$ signal. 
If the magnetic fields have a substantial longitudinal component, a $V/I$ pattern is produced through the Zeeman effect, 
with amplitudes that increase linearly with field strength and that reach roughly $1\%$ in the presence of longitudinal fields 
of $15$~G. Suitably modeling such circular polarization signals requires accounting for the Paschen-Back effect for \gls*{hfs}. 
The magnetograph formula, assuming $L$-$S$ coupling for the Land\'e factors, does not yield a reliable estimate of the 
longitudinal magnetic field from $V/I$. Magnetic fields with transverse components close to $200$~G or larger also produce 
appreciable linear polarization signals due to the Zeeman effect. 

These findings highlight the diagnostic value of spectropolarimetric observations of the Ba~{\sc{ii}} D$_1$.  
However, none of the \gls*{rt} codes currently at our disposal meet all the requirements for a quantitative modeling of the D$_1$ Stokes profiles. 
In particular, it is necessary to account for scattering polarization with \gls*{prd} effects, for the metastable levels and the \gls*{hfs} of the atomic system, 
for magnetic fields in the \gls*{ipb} effect regime, and for the collisional transfer of population and atomic polarization between all levels of the atomic 
system. Fortunately, we expect such a code to be developed in the near future. This would also make it possible to model the D$_1$ together 
with D$_2$, whose large scattering polarization is sensitive to considerably weaker magnetic fields, thus offering complementary information about 
the magnetism in the lower solar chromosphere. \newline

\vspace{0.4cm}
\noindent
 We thank the anonymous reviewer for constructive comments that contributed to improve the manuscript. 
{We} gratefully acknowledge the financial support from the European Research Council (ERC) under the European Union’s Horizon 
2020 research and innovation program (ERC Advanced grant agreement No. 742265). We also acknowledge support from the 
Agencia Estatal de Investigaci\'on del Ministerio de Ciencia, Innovación y Universidades (MCIU/AEI) under grant 
``Polarimetric Inference of Magnetic Fields'' and the European Regional Development Fund (ERDF) with reference 
PID2022-136563NB-I00/10.13039/501100011033. T.P.A.'s participation in the publication is part of the Project 
RYC2021-034006-I, funded by MICIN/AEI/10.13039/501100011033, and the European Union “NextGenerationEU”/RTRP. 
\appendix

\twocolumngrid

\section{Atomic quantities}
\label{sec-app::AtomicQuantTables}
The synthetic Stokes profiles presented in Section~\ref{sec::numerical} were obtained through a two-step calculation 
in a similar manner as was done in \cite{AlsinaBallester+22}. The purpose of the first (or preliminary) step is to provide 
a set of quantities that are required as input in the second step, in which a two-term atomic model is considered. Such 
quantities include the rates of elastic and inelastic collisions, continuum quantities (the thermal emissivity, absorption coefficient, 
and scattering cross-section), and the population of the lower term. They are computed while solving the non-\gls*{lte} 
\gls*{rt} problem considering an atomic model with more levels than in the two-term system. For the sake of reducing 
the computational cost, the problem was solved in the case of unpolarized radiation, using the RH code of \cite{Uitenbroek01}. 
The considered multilevel atomic system consists of five levels for Ba~{\sc{ii}} -- namely the ground level, the two metastable levels 
discussed in the main text, and the upper levels of the D$_1$ and D$_2$ lines -- and of the ground level of Ba~{\sc{iii}}. 
It thus includes $5$ continuum transitions and $5$ line transitions. \gls*{prd} effects are taken into account for the D$_1$ and D$_2$ 
lines but not for the other transitions. We consider this approximation to be suitable for accurately computing the 
aforementioned quantities. The inelastic collisions (those that induce transitions between different terms) were computed 
taking into account only the contribution from free electrons, following \cite{Seaton62}. 
We stored the rate of collisions that couples the upper level of the D$_2$ line and the ground level, and set it equal to 
the broadening rate due to inelastic collisions $\Gamma_I$, to be used in the second step.   
Regarding the rate of elastic collisions (those that induce transitions between states that belong to the same term), 
the quadratic Stark effect contribution from free electrons and singly charged ions was computed following \cite{Traving60} 
and the van der Waals contribution from neutral hydrogen and neutral helium was computed according to \cite{Unsold55}.  
The resulting broadening rate for the upper level of the D$_2$ line was stored as the $\Gamma_E$ broadening rate, to be used 
in the second step. 
 
The second step of the calculation yields the synthetic Stokes profiles of the D$_1$ and D$_2$ lines shown in 
Section~\ref{sec::numerical}. Such profiles are obtained by solving the non-\gls*{lte} \gls*{rt} problem in the polarized 
case using the code described in \cite{AlsinaBallester+22}, which considers a two-term system with \gls*{hfs} but 
cannot account for the metastable levels. The lower term is the $6s\, ^2\mathrm{S}$ ground term, which has a single 
\gls*{fs} level whose energy we take to be zero. The upper term $6p\,^2\mathrm{P}^{\mathrm{o}}$ consists of two \gls*{fs} 
levels: the upper level of the D$_1$ line, with $J = 1/2$ and an energy of $20261.561$ cm$^{-1}$, and the upper level of 
the D$_2$ line, with $J = 3/2$ and an energy of $21952.404$ cm$^{-1}$. 
These energies were taken from the NIST database \citep{NIST_ASD}. In this framework, all transitions are assumed to have 
the same line broadening\footnote{We chose values for the $\Gamma_E$ and $\Gamma_I$ broadenings that correspond to 
the D$_2$ line rather than to D$_1$ or to a weighted mean of the two. Nevertheless, we verified that this choice has 
no appreciable impact on the D$_1$ profile and only a very modest one in the core of the D$_2$ linear polarization profile, 
both when considering the FAL-C and the FAL-P models.}; in addition to the collisional contributions discussed above, this 
broadening has a radiative contribution $\Gamma_R$, which corresponds to the Einstein coefficient for spontaneous 
emission of the term. Because the D lines share the same lower level, the Einstein coefficient of the two lines are identical 
if one assumes $L$-$S$ coupling \citep[e.g., Section 7.5 of][]{BLandiLandolfi04}. In reality, their experimental values differ 
substantially \citep[{see,} e.g.,][]{NIST_ASD}. We take $\Gamma_R = 1.03\times10^8$ s$^{-1}$, which is the average of 
the Einstein coefficients for the two lines accounting for the statistical weights of the upper level of each line. In the second 
step, we take the damping parameter $a$ that enters the \gls*{rt} coefficients to be 
$a = {(\Gamma_R + \Gamma_E + \Gamma_I)}/{(4\pi\Delta\nu_D)}$, where $\Delta\nu_D$ is the Doppler width in frequency units. 

 \begin{table}[!h]
 \centering
 \caption{\label{tab::Isotope} Isotopic abundances and energy shifts}
 \begin{tabular}{|c|c|c|c|c|} \hline
   &                  &     &    \multicolumn{2}{c}{Isotope shift (cm$^{-1}$)}  \vline \\ \cline{4-5}
   & Abundance ($\%$) & $I$ & $6p\,^2\mathrm{P}^{\mathrm{o}}_{1/2}$ & $6p\,^2\mathrm{P}^{\mathrm{o}}_{3/2}$ \\ \hline 
 $^{130}$~Ba & $0.106$  &   $0$      &   $1.185\!\times\!10^{-2}$  & $1.242\!\times\!10^{-2}$ \\ \hline
 $^{132}$~Ba & $0.101$  &   $0$      &   $9.303\!\times\!10^{-3}$  & $9.837\!\times\!10^{-3}$ \\ \hline 
 $^{134}$~Ba & $2.417$  &   $0$       &   $7.425\!\times\!10^{-3}$  & $7.825\!\times\!10^{-3}$ \\ \hline
 $^{135}$~Ba & $6.592$  &  $3/2$    &   $1.163\!\times\!10^{-2}$  & $1.203\!\times\!10^{-2}$ \\ \hline
 $^{136}$~Ba & $7.854$  &   $0$      &   $5.984\!\times\!10^{-3}$  & $6.234\!\times\!10^{-3}$ \\ \hline
 $^{137}$~Ba & $11.232$ &  $3/2$   &   $9.043\!\times\!10^{-3}$  & $9.036\!\times\!10^{-3}$ \\ \hline 
 $^{138}$~Ba & $71.698$ &   $0$     &   $0.000$           & $0.000$ \\ \hline 
 \end{tabular}
\end{table}
The \gls*{hfs} of the atomic system is also included in the second step, in which we account for the seven stable isotopes of barium. 
Their relative abundance, nuclear spin, and their corresponding isotopic shifts for the upper levels of the D$_1$ and D$_2$ lines 
 are displayed in Table~\ref{tab::Isotope}. The isotopic shifts are given relative to the $^{138}$Ba. 
The quantities were taken from Table~1 of \cite{Belluzzi+07}, who themselves took the shifts from \cite{Wendt+84}, except for
 those for the $^{134}$Ba isotope, which were taken from \cite{Wendt+88}. 
 
\begin{table}[!h]
    \centering
    \caption{\label{tab::HFSCoef} Hyperfine structure coefficients}
    \begin{tabular}{|c|c|c|} \hline
       & $^{135}$Ba & $^{137}$Ba \\ \hline 
    $\mathcal{A}_{6s\,^2\mathrm{S}_{1/2}}$ (cm$^{-1}$) & $1.198\!\times\!10^{-1}$ & $1.341\!\times\!10^{-1}$ \\ \hhline{|=|=|=|} 
    $\mathcal{A}_{6p\,^2\mathrm{P}^{\mathrm{o}}_{1/2}}$ (cm$^{-1}$) & $2.217\!\times\!10^{-2}$ & $2.481\!\times\!10^{-2}$ \\ \hline
    $\mathcal{A}_{6p\,^2\mathrm{P}^{\mathrm{o}}_{3/2}}$ (cm$^{-1}$) & $3.769\!\times\!10^{-3}$ & $4.243\!\times\!10^{-3}$ \\ \hline 
    $\mathcal{B}_{6p\,^2\mathrm{P}^{\mathrm{o}}_{3/2}}$ (cm$^{-1}$)  & $1.968\!\times\!10^{-3}$ & $3.085\!\times\!10^{-3}$ \\ \hhline{|=|=|=|} 
    $\mathcal{A}_{5d\,^2\mathrm{D}_{3/2}}$ (cm$^{-1}$) & $5.567\!\times\!10^{-3}$ & $6.329\!\times\!10^{-3}$ \\ \hline 
    $\mathcal{B}_{5d\,^2\mathrm{D}_{3/2}}$ (cm$^{-1}$) & $9.658\!\times\!10^{-4}$ & $1.486\!\times\!10^{-3}$ \\ \hline 
    $\mathcal{A}_{5d\,^2\mathrm{D}_{5/2}}$ (cm$^{-1}$) & $-3.581\!\times\!10^{-4}$ & $-4.012\!\times\!10^{-4}$ \\ \hline 
    $\mathcal{B}_{5d\,^2\mathrm{D}_{5/2}}$ (cm$^{-1}$) & $1.291\!\times\!10^{-3}$ & $1.986\!\times\!10^{-3}$ \\ \hline     
    \end{tabular}
\end{table}
For the isotopes with nonzero nuclear spin, the energies of the various atomic states depend on the $J$, $F$ and $I$ quantum 
numbers through the magnetic dipole ($\mathcal{A}$) and electric quadrupole ($\mathcal{B}$) \gls*{hfs} coefficients that enter 
the Hamiltonian for \gls*{hfs}. The nonzero values of such coefficients are displayed in the four first rows of Table~\ref{tab::HFSCoef}, 
again taken from Table~1 of \cite{Belluzzi+07}, who themselves took the $\mathcal{A}$ coefficients for the ground level from 
\cite{Becker+81} and the other $\mathcal{A}$ and $\mathcal{B}$ coefficients from \cite{Villemoes+93}. The only nonzero $\mathcal{B}$ 
coefficients correspond to the upper level of the D$_2$ line. Such coefficients were defined according to the American convention 
(the expressions of the elements of the \gls*{hfs} Hamiltonian for such convention can be found, for instance, in Appendix B 
of \citealp{AlsinaBallester+22}). 

\begin{figure}[!h]
 \centering
\includegraphics[width = 0.485\textwidth]{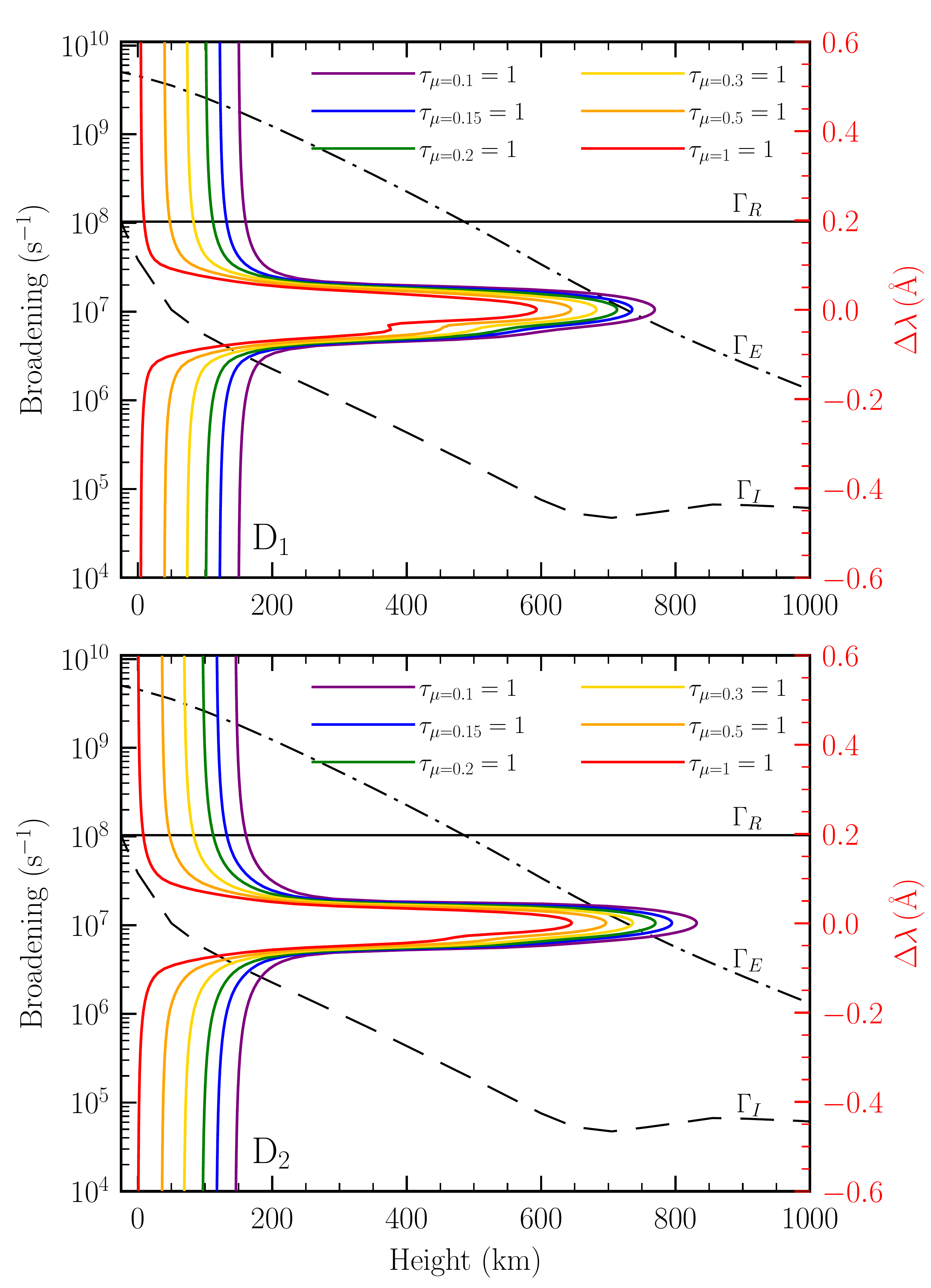}
\caption{Height at which the optical depth is unity in the FAL-C semiempirical atmospheric model as a function of wavelength, 
in a $1.2$~\AA\ spectral range centered on the D$_1$ (upper panel) and D$_2$ (bottom panel) lines, for the \gls*{los}s 
indicated by the colored curves (see the legend). The black curves represent the broadening due to radiative processes 
$\Gamma_R$ (solid curves), inelastic collisions $\Gamma_I$ (dashed curves), and elastic collisions $\Gamma_E$ (dashed-dotted 
curves), as a function of atmospheric height. }
\label{fig-app::FApp1-formation}%
\end{figure} 

The profiles presented in Section~\ref{sec-sub::metastable} were carried out using the HanleRT code, and accounted for the 
metastable levels $5d\,^2\mathrm{D}_{3/2}$ and $5d\,^2\mathrm{D}_{5/2}$. Their \gls*{hfs} coefficients, taken from 
\cite{Silverans+86}, are shown in the four bottom rows of Table~\ref{tab::HFSCoef}. The isotopic shifts for the five most 
abundant isotopes, which were the ones considered in such same calculations, were obtained from \cite{Villemoes+93}. 

\section{Formation Height}
\label{sec-app::Formation}

Figure~\ref{fig-app::FApp1-formation} shows the height in the FAL-C model at which the optical depth $\tau_\nu$ is equal 
to unity. It is shown in two $1.2$~\AA-wide spectral ranges, centered on the D$_1$ (discussed in the main text) and the 
D$_2$ lines (discussed in Appendix~\ref{sec-app::FigsD2}). This height is a proxy for the formation height of the line and 
it is shown for several \gls*{los}s. In a \gls*{1d} atmospheric model, the optical depth is given by 
$\mathrm{d}\tau_\nu = -\eta_I \,\mathrm{d}z/\!\mu$, where $z$ is the atmospheric height. $\eta_I$ is the absorption 
coefficient, which was computed following \cite{AlsinaBallester+22}, taking $\mathcal{N}_\ell$ as obtained in step $1$ 
of the approach described in Section~\ref{sec-sub::form-num-framework}. We note that the line core of the D$_1$ forms 
above the temperature minumum; the heights at which $\tau = 1$ at line center are just below $600$~km in the 
FAL-C model for an \gls*{los} with $\mu = 1$ and above $750$~km for $\mu = 0.1$. The D$_2$ forms at slightly higher 
regions; the height at which its optical depth is equal to unity in the FAL-C model is above $600$~km for $\mu = 1$ and 
above $800$~km for $\mu = 0.1$. The $\Gamma_R$, $\Gamma_E$, and $\Gamma_I$ broadenings, shown in both panels 
as a function of height for reference, were obtained as explained in Appendix~\ref{sec-app::AtomicQuantTables}. 

\section{Analysis of the D${}_2$ Stokes profiles}\label{appC} 
\label{sec-app::FigsD2}
In the main text, we discussed how various features of the atomic system and properties of the solar atmosphere 
impact the intensity and linear polarization pattern of the Ba~{\sc{ii}} D$_1$ line, illustrated by the synthetic profiles 
displayed in Sections~\ref{sec-sub::metastable} and \ref{sec::numerical}. The profiles presented therein were calculated 
considering atomic models that considered not only the D$_1$ line transitions at $4934$~\AA , but also the D$_2$ 
line transitions at $4554$~\AA . In this appendix, we show the synthetic profiles in the spectral range around the 
D$_2$ line instead of D$_1$. Many such profiles were obtained through the same calculations that yielded the profiles 
shown in the main text. The FAL-C atmospheric model was considered for all such calculation. Except where otherwise 
noted, the profiles are shown for an \gls*{los} with $\mu = 0.1$, taking the reference direction for positive Stokes $Q$ 
parallel to the nearest limb. 
\begin{figure}[!h]
  \centering
\includegraphics[width = 0.485\textwidth]{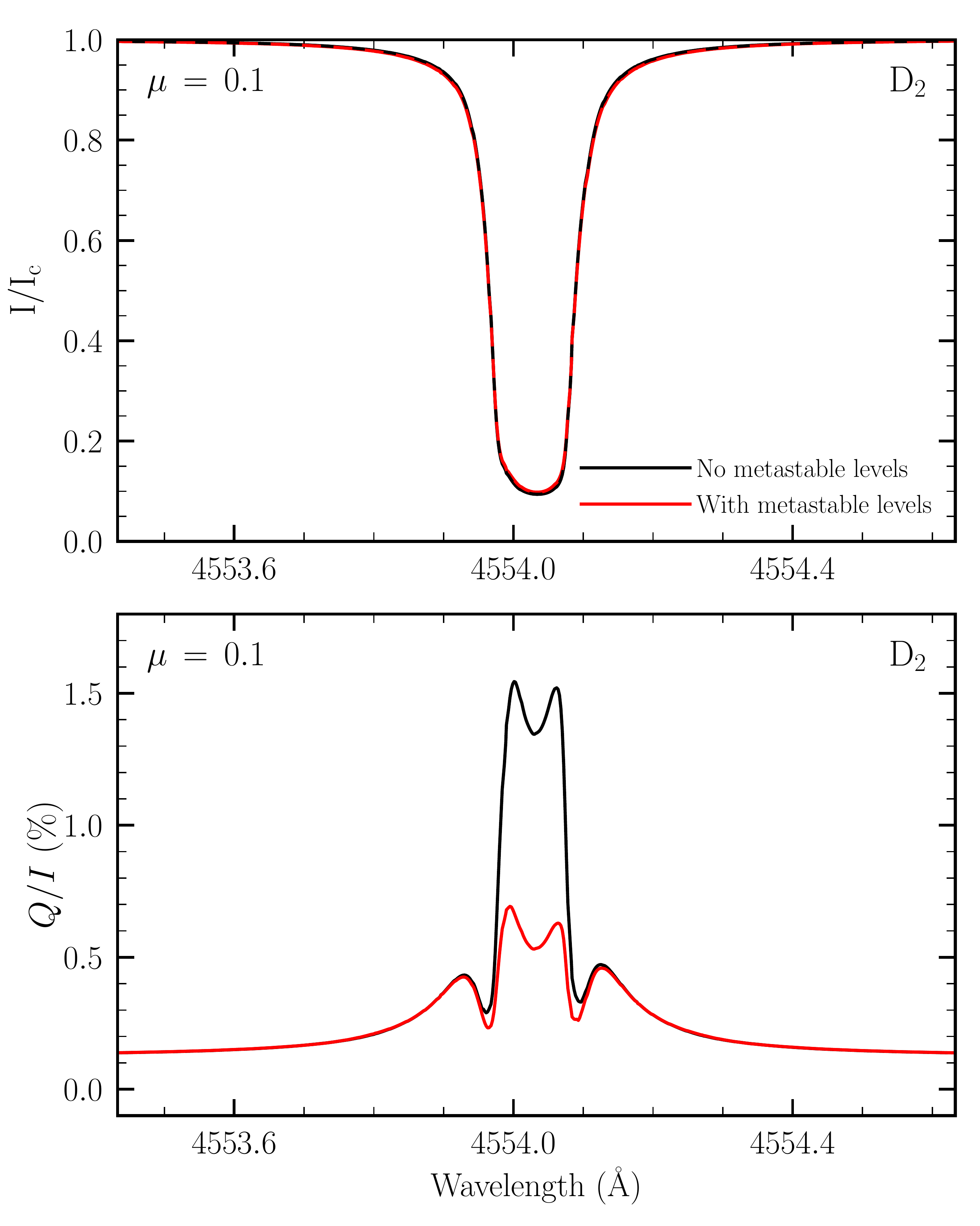}
 \caption{Stokes $I$ ({upper} panel), normalized to the continuum intensity $I_\mathrm{c}$, and $Q/I$ profiles (lower panel) 
 of the D$_2$ line as a function of wavelength (i.e., in a $1.2$~\AA -wide range centered on the D$_2$ line). 
 The synthetic profiles are obtained from calculations using the HanleRT code, accounting for \gls*{prd} effects and taking a line of sight 
 (LOS) with $\mu = 0.1$. The black curves correspond to the calculations considering a five-level atomic model, including the levels 
 belonging to the $6s\,^{2}\mathrm{S}$ ground term, the $6p\,{}^2\mathrm{P}$ term that contains the upper levels of the D lines, 
 and the $5d\,^{2}\mathrm{D}$ metastable term. The red curves correspond to the calculations in which the metastable term is neglected, 
 for which a three-level atomic model is thus considered. In both cases, the \gls*{hfs} of the atomic system is taken into account. 
 The population of the ground level, $\mathcal{N}_\ell$, is kept fixed during the iterative calculation.} 
  \label{fig-app::Fapp4-HanleRT-D2-metast}
\end{figure}

The D$_2$ Stokes profiles show in Figure~\ref{fig-app::Fapp4-HanleRT-D2-metast} were obtained simultaneously with 
those presented in Figure~\ref{fig::F2-HanleRT-D1-metast}. Thus, the figure shows the profiles obtained with HanleRT 
using the atomic models discussed in Section~\ref{sec-sub::metastable}, taking into account the \gls*{hfs} of the odd 
isotopes for all considered levels, both accounting for the metastable levels and neglecting them. Although the 
$5d\,^2\mathrm{D}$ levels do not substantially change the D$_2$ intensity profile, they have a crucial impact on its linear 
polarization, decreasing its line-core amplitude by roughly $60\%$. The depolarization due to the metastable levels is far 
greater than the one reported for the D$_1$ line, and also leads to a far more apparent change in the shape of its $Q/I$ pattern. 
For this reason, we deem the calculations presented in Section~\ref{sec::numerical}, for which the metastable levels were not 
included, to be less reliable for the D$_2$ line than for D$_1$. Despite this, they may still provide some insights into the sensitivity 
of this line to the \gls*{hfs} or to the magnetic field. We also verified that the dips found in the line center of the $Q/I$ profile are a 
consequence of making the \gls*{aa} approximation, implying that a strictly correct modeling of the D$_2$ line should be 
carried out through a fully angle-dependent calculation, in contrast to the case of the D$_1$ line. 

\begin{figure}[!h]
  \centering
  \includegraphics[width = 0.485\textwidth]{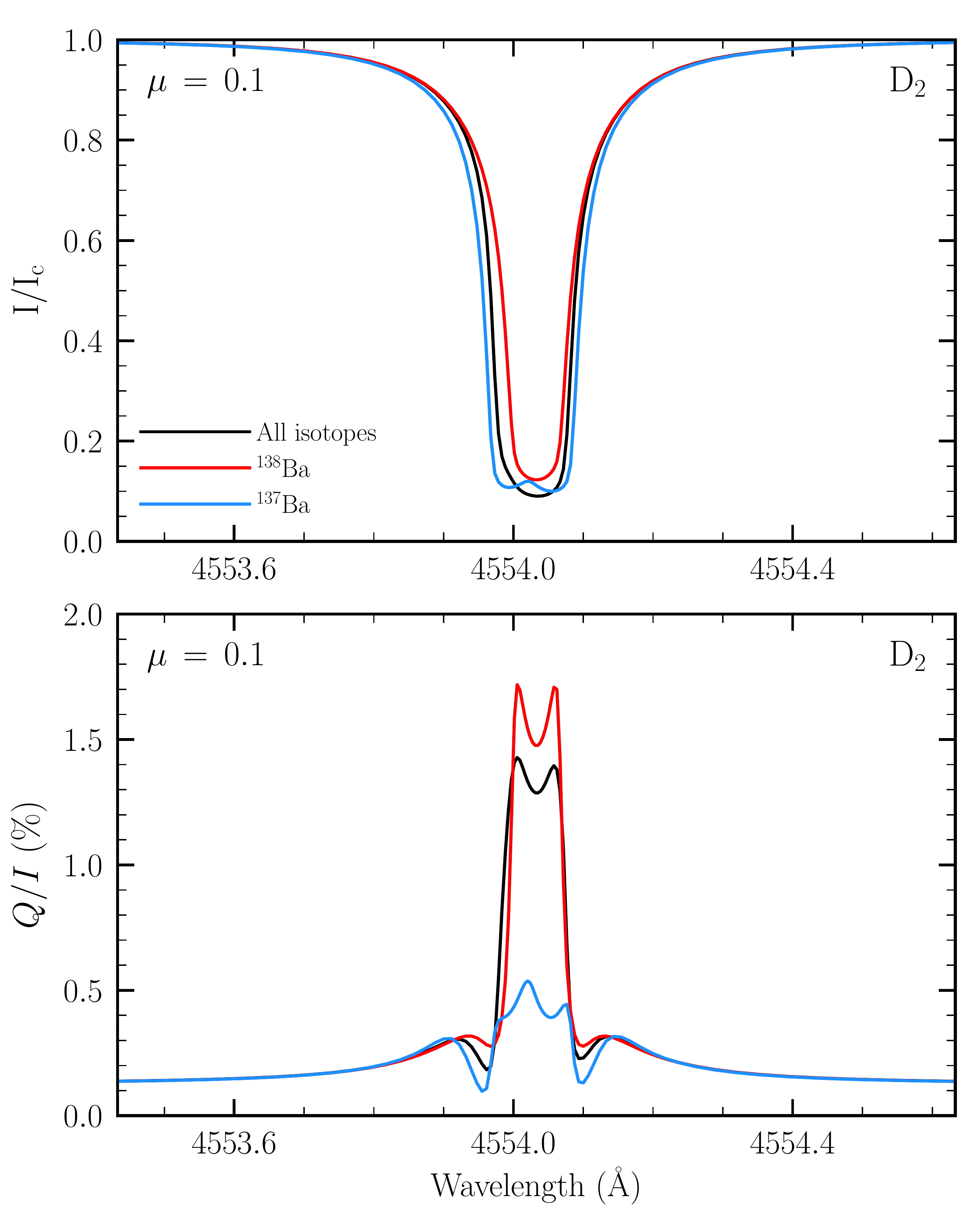}
  \caption{Stokes $I$ (upper panel), normalized to $I_\mathrm{c}$, and $Q/I$ profiles (lower panel) of the D$_2$ line as a 
  function of wavelength. The profiles calculated considering different isotopes of barium are shown with different colored curves. 
  The black curve corresponds to the case in which all seven stable isotopes are considered with their corresponding abundances, 
  whereas the red and blue curves correspond to the cases in which only the $^{138}$Ba and only the $^{137}$Ba isotopes were 
  considered, respectively.} 
  \label{fig-app::FApp5-D1D2HFS}
\end{figure}
In the rest of this appendix, we present profiles using the numerical code discussed in \cite{AlsinaBallester+22}, most of them 
being analogous to those presented in Section~\ref{sec::numerical} for the D$_1$ line. First, we study the impact of the \gls*{hfs} of the 
odd isotopes but, unlike in Section~\ref{sec-sub::num-hfs}, we do not compare the intensity and linear polarization profiles obtained by 
accounting for the \gls*{hfs} splitting of different \gls*{fs} levels and neglecting it. Instead, as shown in Figure~\ref{fig-app::FApp5-D1D2HFS}, we 
compare the profiles obtained considering all seven stable isotopes with their corresponding abundances (see Appendix~\ref{sec-app::AtomicQuantTables}), 
considering only the $^{137}$Ba isotope, which has nuclear spin $I = 3/2$ and \gls*{hfs}, and only $^{138}$Ba, for which $I = 0$ and 
thus has no \gls*{hfs}. The inclusion of isotopes with \gls*{hfs} leads to a broadening of the absorption profile in intensity, 
much like what was reported in the D$_1$ line. 
A comparison between the linear polarization profiles obtained considering isotopes with $I = 0$ and $3/2$ reveals 
that the very large \gls*{hfs} of barium depolarizes its line-core $Q/I$ by almost a factor $4$. This is consistent with the theoretical 
depolarization when the quantum interference between different $F$-levels of the upper term is negligible 
\citep[e.g., Section~10.22 of][]{BLandiLandolfi04}. The full scattering polarization amplitude is thus mainly produced by the even 
isotopes (which have no \gls*{hfs}). This clearly contrasts with the linear polarization pattern of the D$_1$ line, which is a consequence of 
the wavelength separation between the \gls*{hfs} components of the odd isotopes. 

\begin{figure}[!h]
  \centering
  \includegraphics[width = 0.485\textwidth]{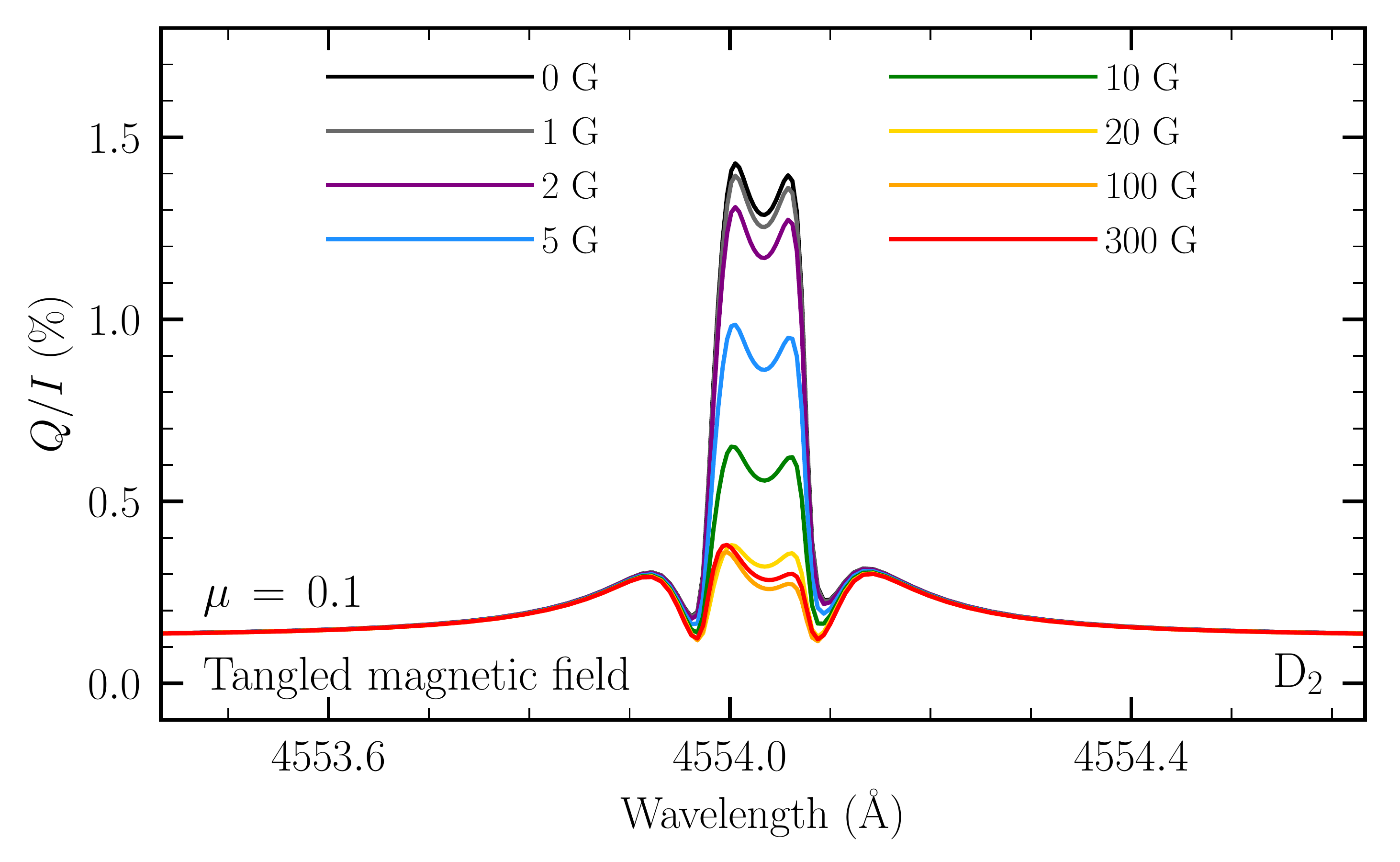}
  \caption{Synthetic $Q/I$ profiles of the D$_2$ line as a function of wavelength, obtained in 
   the presence of tangled magnetic fields of the various strengths indicated in the legend.}  
  \label{fig-app::FigApp7-D2Microturb}%
\end{figure} 
In this Appendix, we also show the D$_2$ profiles obtained in the presence of both tangled and deterministic magnetic fields, as in 
Section~\ref{sec-sub::magnetic}, but considering different field strengths in the range between $0$ and $300$~G. 
The intensity profiles are not appreciably affected by fields of such strengths and thus they are not shown here. The sensitivity of the D$_2$ 
linear polarization to tangled magnetic fields (see Section~\ref{sec-subsub::isotropic}) is illustrated in Figure~\ref{fig-app::FigApp7-D2Microturb}. 
Such fields preserve the axial symmetry of the problem and thus no $U/I$ or $V/I$ signal is produced. For this line, a depolarization 
is clearly appreciable in $Q/I$ for fields as weak as $2$~G -- considerably lower than those required to modify the D$_1$ signal. 
We note that, in contrast to the D$_1$ line, most of the contribution to the D$_2$ scattering polarization comes from the roughly 
$82\%$ of isotopes without \gls*{hfs}. Neglecting \gls*{hfs}, the magnetic field at which the Zeeman splitting of the upper 
level of D$_2$ is equal its the natural width \citep[i.e., the Hanle critical field; see e.g.,][]{BStenflo94} is approximately $9$~G. 
This is fully consistent with the behavior displayed in Figure~\ref{fig-app::FigApp7-D2Microturb}. 
For magnetic fields stronger than $100$~G, the linear polarization amplitude slightly increases, due to the \gls*{hfs} 
of the odd isotopes, until reaching saturation \citep[for a further discussion on this enhancement, see Section~10.22 of][]{BLandiLandolfi04}. 
 
\begin{figure}[!h]
  \centering 
 \includegraphics[width = 0.485\textwidth]{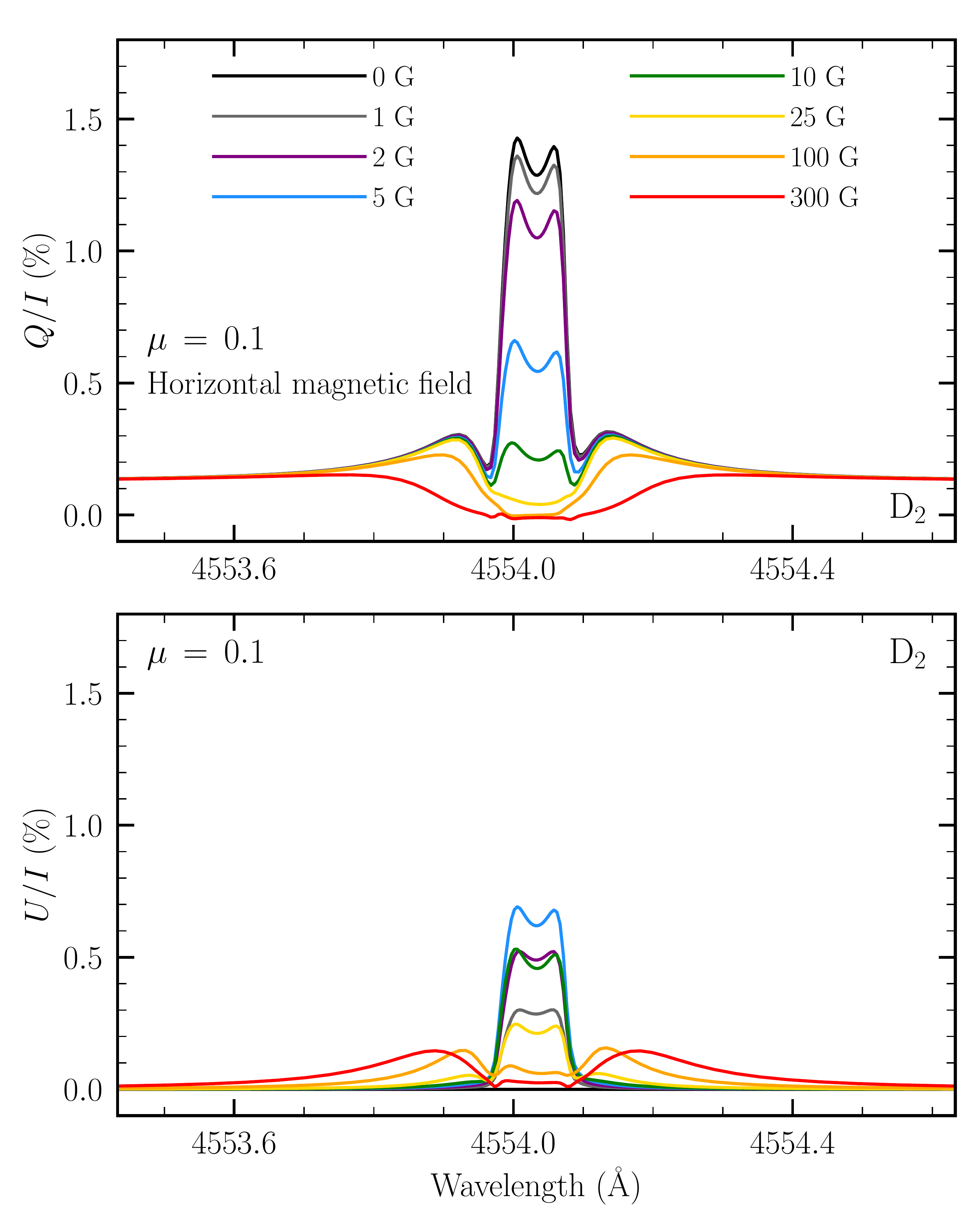} 
  \caption{Synthetic $Q/I$ (top panel) and $U/I$ profiles {(bottom panel)} of the D$_2$ line as a function of wavelength, obtained 
  in the presence of deterministic horizontal fields contained in the plane defined by the local vertical and the \gls*{los}. The various 
  colored curves represent the calculations carried in the presence of magnetic fields of the strengths indicated in the legend.} 
  \label{fig-app::D2-QU-Horiz}%
\end{figure} 
The synthetic $Q/I$ and $U/I$ profiles of the D$_2$ line, obtained in the presence of deterministic horizontal magnetic fields, 
contained in the plane defined by the local vertical and the \gls*{los}, are shown in Figure~\ref{fig-app::D2-QU-Horiz}. All 
the considered fields have a positive projection onto the \gls*{los} and the various curves indicate the same field strengths as in
Figure~\ref{fig-app::FigApp7-D2Microturb}. In the case of horizontal magnetic fields, we observe a stronger depolarization in 
the line core for a given field strength and a near-zero $Q/I$ value for fields larger than $100$~G, in contrast to the substantial 
saturation value found for tangled fields. Moreover, horizontal fields induce a rotation of the plane of linear polarization as the 
quantum interference between nearby $f$-states is modified (i.e., Hanle rotation). A maximum in the $U/I$ amplitude is reached 
for field strengths close to $5$~G; as the field becomes stronger, the interference between $f$ states becomes smaller and the 
linear polarization fraction decreases. We see no indication of the loop-like behavior in the polarization diagram (i.e., increases in 
the $U/I$ amplitude with field strength after having reached a first maximum) that is found in D$_2$ lines of K~{\sc{i}} 
\citep[see][]{AlsinaBallester22} and of Na~{\sc{i}} \citep[see Section 10.22 of][]{BLandiLandolfi04}. We attribute the absence of such 
loops to the fact that the \gls*{hfs} splitting of the Ba~{\sc{ii}} isotopes with nuclear spin (required for the loop-like behavior) is large 
enough that no level crossings are reached until field strengths of roughly $50$~G are considered, at which point the line is almost depolarized. 

\begin{figure}[!h]
  \centering 
  \includegraphics[width = 0.485\textwidth]{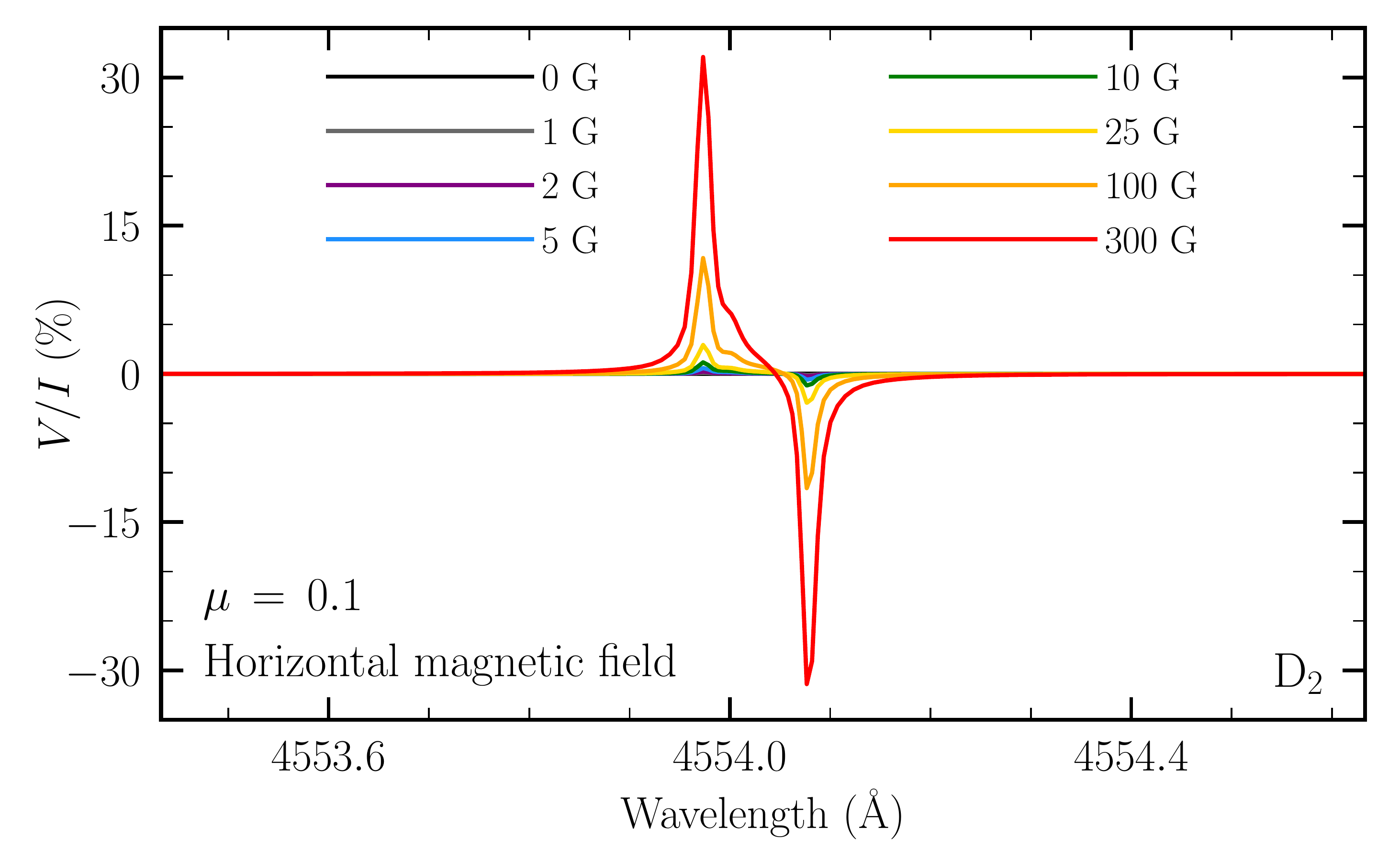}
  \caption{Synthetic $V/I$ profiles as a function of wavelength, in the spectral range centered on D$_2$, obtained in the presence of 
  the same horizontal fields as in the previous figure.} 
  \label{fig-app::D2-V-Horiz}%
\end{figure} 
The considered horizontal fields present a substantial longitudinal component for an \gls*{los} with $\mu = 0.1$ and, thus, a clear 
$V/I$ pattern is produced, as shown in Figure~\ref{fig-app::D2-V-Horiz}. The amplitude of the signal increases linearly with the field 
strength, and reaches roughly $1\%$ for $10$~G fields. Like in the case of the D$_1$ line, reproducing the shape of the $V/I$ pattern 
requires accounting for the \gls*{hfs} of the barium atoms. We do not find two distinct blue peaks in the D$_2$ line, but the \gls*{hfs} 
splitting does contribute to broadening it appreciably. 
\begin{figure}[!h]
  \centering
  \includegraphics[width = 0.485\textwidth]{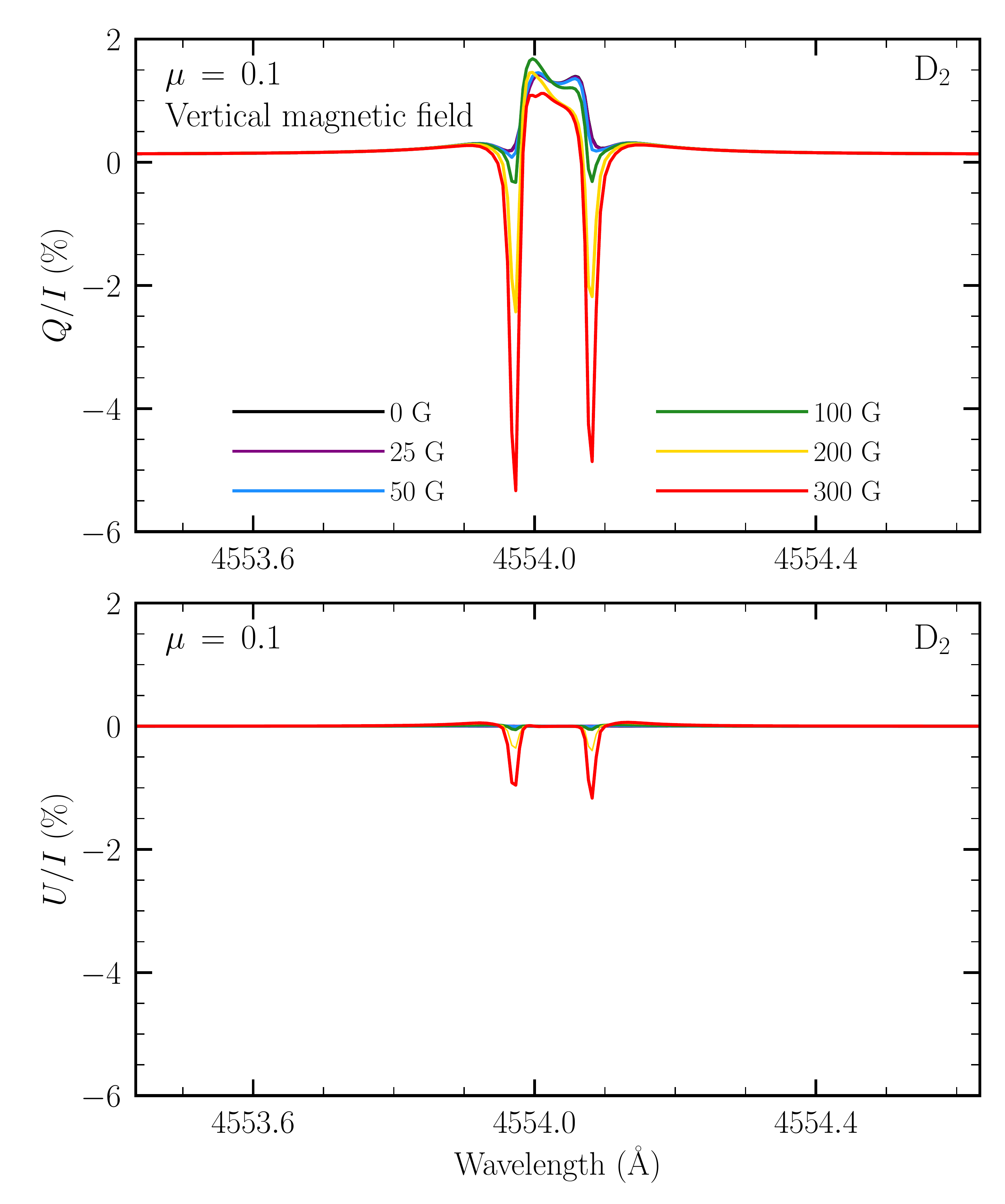} 
  \caption{Synthetic $Q/I$ (top panel) and $U/I$ (bottom panel) profiles of the D$_2$ line as a function of wavelength, in the presence of 
  a vertical magnetic field. The colored curves represent the results of calculations carried in the presence of magnetic fields of the strengths 
  indicated in the legend.}
  \label{fig-app::D2-QU-Vert}
\end{figure}  
\begin{figure}[!h]
  \centering
  \includegraphics[width = 0.485\textwidth]{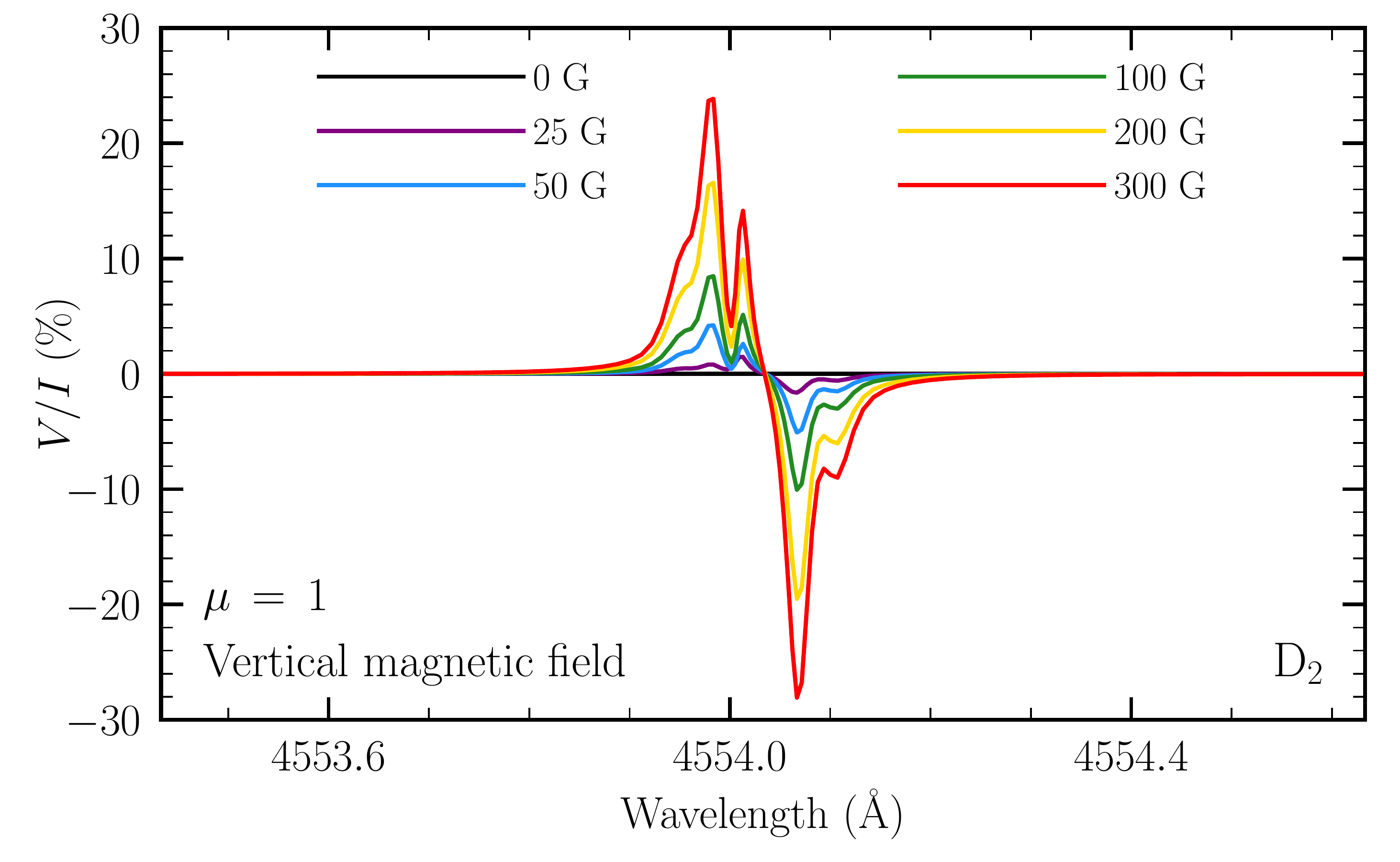}
  \caption{Synthetic $V/I$ profiles as a function of wavelength, in the spectral range centered on D$_2$, obtained in the presence of a 
  vertical magnetic field, for an \gls*{los} with $\mu = 1$. The colored curves represent the $V/I$ profiles obtained for the field strengths
   indicated in the legend.}   
  \label{fig-app::D2-V-Vert}%
\end{figure} 

We also considered the case of a vertical magnetic field, and the resulting linear polarization patterns for an \gls*{los} with $\mu = 0.1$ are 
shown in Figure~\ref{fig-app::D2-QU-Vert}. Vertical magnetic fields only modify the quantum interference between states with the 
same quantum number $f$, and thus its impact on the scattering polarization is much more modest than that of tangled or horizontal magnetic fields. 
Magnetic fields of about $100$~G or stronger have a large transverse component and introduce a further linear polarization signal 
due to the Zeeman effect, whose clearest signature in the figure are the negative peaks at either side of the line core, with amplitudes that 
are noticeably larger than the signals found for the D$_1$ line for the same geometry and field strengths (see Figure~\ref{fig::D1-QU-Vert}). 
Moreover, vertical magnetic fields also give rise to a circular polarization pattern in the D$_2$ line due to the Zeeman effect, 
as is shown in Figure~\ref{fig-app::D2-V-Vert}. An \gls*{los} with $\mu = 1$  was selected for this figure in order to consider a completely longitudinal magnetic field. 
For this geometry, we do find a two-peak structure to the blue of the line center. In contrast to the D$_1$ $V/I$ profile, the bluemost 
peak of the D$_2$ presents a larger amplitude than the one closer to line center. The peak to the red of the line center is much sharper 
for the D$_2$ line than for D$_1$, but it also presents a secondary lobe. We also verified that the amplitude of the D$_2$ $V/I$ is 
also substantially underestimated when neglecting the elements of the magnetic Hamiltonian that are off-diagonal in $F$ (making the linear Zeeman
approximation, see Section~\ref{sec-subsub::horiz}). Likewise, we also find the magnetograph formula to be unsuitable for reproducing 
the D$_2$ $V/I$ pattern while considering the $L$-$S$ coupling scheme for the effective Land\'e factor, although the error incurred is not as 
large as for the D$_1$ line. 

\bibliographystyle{aasjournal}
\bibliography{ms}

\end{document}